\documentclass[letterpaper, 10 pt, conference]{ieeeconf}  % Comment this line out
\IEEEoverridecommandlockouts                              % This command is only
\usepackage{graphics} % for pdf, bitmapped graphics files
\usepackage{epsfig} % for postscript graphics files
\usepackage{mathptmx} % assumes new font selection scheme installed
\usepackage{times} % assumes new font selection scheme installed
\usepackage{amsmath} % assumes amsmath package installed
\usepackage{amssymb}  % assumes amsmath package installed
\usepackage{booktabs}
\usepackage{algorithm}
\usepackage{algpseudocode}

\usepackage{soul}
\usepackage{subcaption}
\usepackage[utf8]{inputenc}
\usepackage{textgreek}
\soulregister\cite7
\soulregister\ref7
\soulregister\eqref7
\usepackage{xcolor}
\usepackage{float}
\usepackage{stfloats}
\usepackage{threeparttable}
\usepackage{hyperref}
\usepackage{textcomp}
\usepackage{afterpage}
\usepackage{multirow}
\usepackage{placeins} 

\title{\LARGE \bf
EMind: A Foundation Model for Multi-task Electromagnetic Signals Understanding
}

\author{
Luqing Luo\textsuperscript{1}$^\dagger$, Wenjin Gui\textsuperscript{4}$^\dagger$, 
Yunfei Liu\textsuperscript{3}, Ziyue Zhuang\textsuperscript{4}, Yunxi Zhang\textsuperscript{4},\\
Fengxiang Wang\textsuperscript{5}, Zonghao Guo\textsuperscript{2}, Qirui Zhao\textsuperscript{7}, Zizhi Ma\textsuperscript{6},
Xinzhu Liu\textsuperscript{3}, Hanxiang He\textsuperscript{4},\\
Jinhai Li\textsuperscript{1}, Xin Qiu\textsuperscript{1}, Wupeng Xie\textsuperscript{3}*, 
Yangang Sun\textsuperscript{2}*\\[1ex]
\textsuperscript{1}Institute of Microelectronics of the Chinese Academy of Sciences, China;
\textsuperscript{2}Tsinghua University, China \\
\textsuperscript{3}Artificial Intelligence Institute of China Electronics Technology Group Corporation, China \\
\textsuperscript{4}Beijing Institute of Technology, China;
\textsuperscript{5}National University of Defense Technology, China \\
\textsuperscript{6}Nankai University, China;
\textsuperscript{7}Northeastern University, China \\
$^\dagger$\small These authors contributed equally to this work.
}

\usepackage[T1]{fontenc}
\usepackage{graphicx}
\usepackage{caption}

\begin{document}

\maketitle
\thispagestyle{empty}
\pagestyle{empty}

%%%%%%%%%%%%%%%%%%%%%%%%%%%%%%%%%%%%%%%%%%%%%%%%%%%%%%%%%%%%%%%%%%%%%%%%%%%%%%%%
\begin{abstract}
% 
% 电磁信号的深度理解是实现动态频谱管理、智能交通与无人驾驶、无人机与机器人感知等多领域核心能力的基础。然而，当前该领域面临诸多挑战。首先，电磁信号作为一种独特模态，与文本和图像有显著差异，具有高度异构性、强背景噪声以及复杂的时频联合结构特征，使得现有的通用模型架构难以直接适用。此外，电磁信号通信与理解任务种类繁多，现有方法缺乏足够的跨任务泛化能力和迁移效率，在高质量、大规模电磁信号数据稀缺的背景下，难以构建出一个泛化性足够强的多任务学习框架。
% 为突破这些瓶颈，我们提出了一个为电磁信号量身定制的专用领域基础模型——EMind，弥合了大规模预训练范式与电磁信号模态之间的差距。具体而言，我们构建了首个统一、标准化、涵盖多信号类别及多任务的已知规模最大电磁信号数据库。结合电磁信号的物理特性，我们提出了一种长度自适应的多信号打包方法和一种硬件感知的训练策略，实现了对多源异构、信号长度不一的数据的高效利用与特征表示学习。我们在一个框架内验证了多种下游任务的有效性和良好泛化能力，标志着电磁智能从传统专用模型向统一理解框架的关键跃迁。代码可在以下地址获取：https://github.com/GabrielleTse/EMind.
Deep understanding of electromagnetic signals is fundamental to dynamic spectrum management, intelligent transportation, autonomous driving and unmanned vehicle perception. The field faces challenges because electromagnetic signals differ greatly from text and images, showing high heterogeneity, strong background noise and complex joint time frequency structure, which prevents existing general models from direct use. Electromagnetic communication and sensing tasks are diverse, current methods lack cross task generalization and transfer efficiency, and the scarcity of large high quality datasets blocks the creation of a truly general multitask learning framework. To overcome these issue, we introduce EMind, an electromagnetic signals foundation model that bridges large scale pretraining and the unique nature of this modality. We build the first unified and largest standardized electromagnetic signal dataset covering multiple signal types and tasks. By exploiting the physical properties of electromagnetic signals, we devise a length adaptive multi-signal packing method and a hardware-aware training strategy that enable efficient use and representation learning from heterogeneous multi-source signals. Experiments show that EMind achieves strong performance and broad generalization across many downstream tasks, moving decisively from task specific models to a unified framework for electromagnetic intelligence. The code is available at: https://github.com/GabrielleTse/EMind.
\end{abstract}
%%%%%%%%%%%%%%%%%%%%%%%%%%%%%%%%%%%%%%%%%%%%%%%%%%%%%%%%%%%%%%%%%%%%%%%%%%%%%%%%

\section{Introduction}
\begin{figure*}[t]
    \centering
    \includegraphics[width=0.95\linewidth]{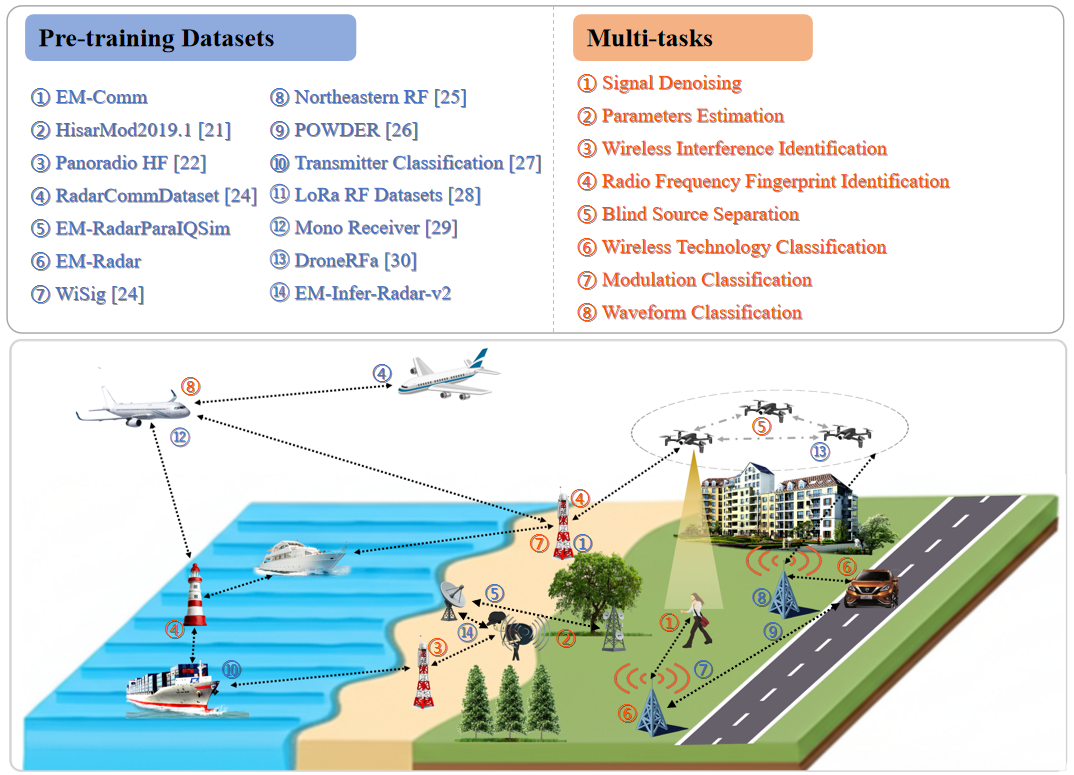}
    \caption{EMind. A foundation model for electromagnetic signals capable of multitask learning, including signal denoising, parameter estimation, modulation classification and interference identification, applied across diverse fields such as communication, navigation, and security.}
    \label{fig:big_pic}
\end{figure*}
%
% 我们生活在一个被电磁波环绕的世界，这些看不见却无处不在的信号承载着丰富的信息，广泛应用于通信、导航、环境感知等关键领域。对电磁信号的理解任务，涵盖信号属性识别、目标特征辨别、干扰判别与波形建模等多个层面，支撑动态频谱接入和智能感知等先进应用，是实现复杂电磁环境下智能感知与决策的核心能力，推动智能交通、智慧城市、无人驾驶和物联网等场景的发展。
Living in a world surrounded by electromagnetic waves, these invisible yet omnipresent signals carry abundant information and are widely applied in critical fields such as communication, navigation, and environmental perception. The tasks of understanding electromagnetic (EM) signals encompass various aspects, including signal property recognition, target feature identification, interference detection, and waveform reconstruction. These tasks support advanced applications such as dynamic spectrum access and intelligent perception, forming the core capability for intelligent sensing and decision-making in complex electromagnetic environments. They also drive the development of scenarios like intelligent transportation, smart cities, autonomous driving, and the Internet of Things (IoT).
\begin{minipage}{\linewidth}
  \centering
  \includegraphics[width=1.01\linewidth]{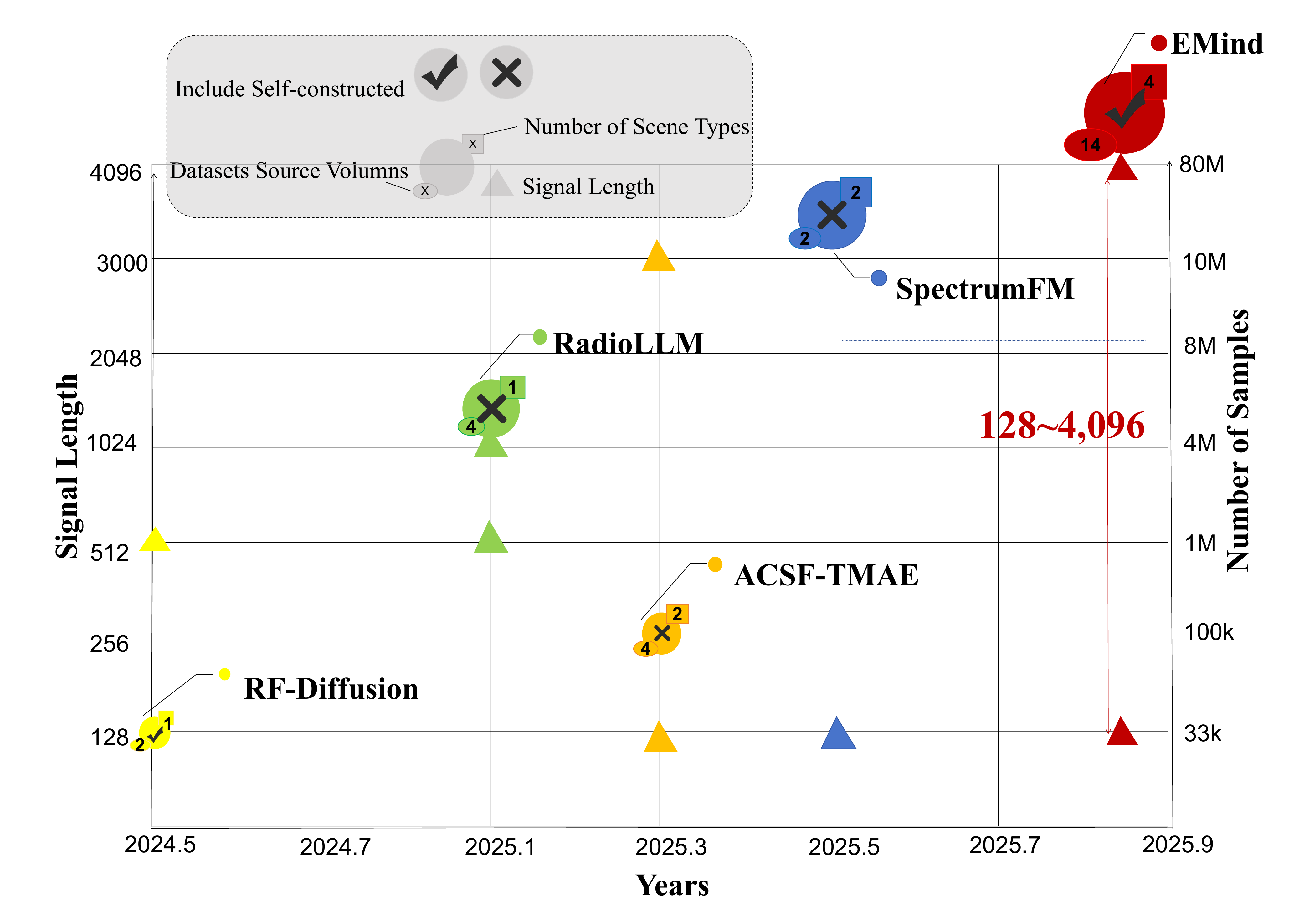}
  \captionof{figure}{Comparison of EM signals pre-train datasets: number of samples, data source volume, scene types and signal lengths.}
  \label{fig:comp}
\end{minipage}
%
% 近年来，深度学习在电磁信号通信与感知领域展现出显著优势，推动了从传统规则驱动方法和特征工程向数据驱动范式的转变~\cite{hao2023contrastive, zhang2023self}。然而，目前大多数方法依然基于专用任务模型，面临着泛化能力不足和适应新任务的高成本等挑战，因此迫切需要一个通用的基础模型，以提升电磁信号在多任务中的理解能力。
In recent years, deep learning has demonstrated significant advantages in the field of EM signal communication and sensing, driving a shift from traditional rule-based methods and feature engineering toward a data-driven paradigm~\cite{hao2023contrastive, zhang2023self}. However, most existing methods still rely on specialized task-specific models, confronting challenges such as insufficient generalization capabilities and high costs to adapt to new tasks. Therefore, there is an urgent need for a generalized foundation model to enhance the understanding of EM signals across multiple tasks. 
% 为支持预训练-微调范式在电磁信号模态上的验证，我们迫切需要构建大规模数据集。然而，由于大多数电磁信号数据来自非合作场景，协议标准通常不公开，甚至存在加密保护，导致现有数据稀缺、零散且非规范化，高质量数据的获取极为困难。因此，我们系统整理并构建了迄今为止规模最大统一、标准化、涵盖多类信号与任务的电磁信号数据集，以支持大规模预训练，如图~\ref{fig:comp}所示，为通用特征表示学习奠定了坚实的数据基础。
To support the validation of the pre-training and fine-tuning paradigm on EM signal modalities, it is imperative to construct large-scale datasets. Nevertheless, since most EM signal data originate from non-cooperative scenarios, the protocol standards are generally not public and even subject to encryption protection. This results in existing data being scarce, fragmented, and non-standardized, making the acquisition of high-quality data extremely difficult. Therefore, we have systematically organized and constructed to date the largest unified, standardized EM signal dataset covering multiple signal types and tasks to support large-scale pre-training. As shown in Figure~\ref{fig:comp}, this dataset provides a firm basis for universal feature representation learning.

% 尽管近年来已有研究尝试将自然语言处理（NLP）、计算机视觉（CV）及通用时间序列（TS）建模中的基础模型范式迁移到电磁信号领域，电磁信号作为一种独特模态，在时频联合域中同时呈现稀疏性与连续性并存、非离散化的结构特性，既不同于离散的文本序列，也迥异于二维图像数据。此外，电磁信号来源多样、类型复杂，具有高度异构性，直接沿用 NLP、CV 或 TS 的通用模型仍面临诸多挑战。因此，亟需构建专门针对电磁信号物理属性并适配其特性的基础模型架构。基于此，我们提出面向电磁信号的基座模型——EMind，并为其设计了 IQ 信号专用架构。EMind 通过 length-adaptive multi-signal packing 与 per-sample masking 机制，实现对电磁信号模态的高效建模与特征提取。
While recent efforts have transplanted the foundational-model paradigm from Natural Language Processing (NLP), Computer Vision (CV)~\cite{jing2020self}, and general Time Series (TS) modeling to the field of EM signals, the electromagnetic modality remains fundamentally distinct. RF waveforms exhibit simultaneous sparsity and continuity in the joint time–frequency domain, are inherently non-discrete, and are sourced from a highly heterogeneous space—properties that diverge sharply from tokenized text or 2-D imagery. Naïvely repurposing existing architectures therefore incurs substantial representational misalignment. To close this gap, we introduce EMind, a foundation model for multi-task learning of communication and sensing on EM signals. EMind’s architecture integrates length-adaptive multi-signal packing with per-sample masking, enabling efficient modeling and robust feature extraction tailored to the physical characteristics of electromagnetic data.

% 在统一数据基座与模型架构之上，我们系统性地验证了“预训练–微调”范式于电磁智能的普适性。具体而言，单一主干网络即可覆盖自动调制分类（AMC）、雷达波形识别（RWC）、雷达参数估计（RPE）、无线干扰判别（WII）、射频指纹提取（RFFI）等判别任务，并进一步延伸至盲源分离（BSS）与信号去噪（SD）这类高复杂度的生成式重建任务。尤为关键的是，我们将轻量自动编码器接入基座模型，实验结果表明，在基于重建的生成场景下，模型仍能受益于预训练权重的初始化，获得可观的性能提升。这一观察不仅丰富了预训练–微调范式在电磁任务中的适用谱系，也为后续研究提供了新的实证线索，推动电磁智能逐步由专用模型迈向面向通用理解的新阶段。总之，本文的主要贡献如下：
Building upon a unified data foundation and model architecture, we systematically validate the generalizability of the pre-training–fine-tuning paradigm in electromagnetic intelligence. Specifically, a single backbone network is shown to accommodate discriminative tasks such as Automatic Modulation Classification (AMC), Radar Waveform Classification (RWC), Radar Parameter Estimation (RPE), Wireless Interference Identification (WII), and Radio-Frequency Fingerprinting Identification (RFFI), while further extending to high-complexity generative reconstruction tasks including Blind Source Separation (BSS) and Signal Denoising (SD). Crucially, by integrating auto-encoders (AE) into the base model, we demonstrate that in reconstruction-oriented generative scenarios, the model continues to benefit from initialization with pre-trained weights, yielding substantial performance improvements. This observation not only enriches the applicability spectrum of the pre-training–fine-tuning paradigm across electromagnetic signal tasks but also provides new empirical evidence that advances electromagnetic intelligence from specialized models toward a new stage of generalized understanding. In summary, the following main contributions are presented in this paper:

% 我们提出了一个专为电磁信号量身定制的基础模型，弥合了大规模预训练范式与电磁信号模态之间的差距，为构建具有卓越表征能力的电磁信号基础模型奠定了坚实基础。
% 我们设计了适用于电磁模态的网络架构，针对电磁信号序列长度不一、时域稀疏性与连续性并存等模态特征，提出了信号离散化机制、可变长度输入处理方法，以及面向大规模异构数据的高效训练优化策略。
% 我们构建了迄今为止最大规模的电磁信号数据集，涵盖多类信号、多种属性标签和多种下游任务，并以统一标准化与高效存储方案打造坚实数据底座，支撑大规模预训练与通用特征表示学习。
% EMind统一验证预训练–微调在电磁任务的普适性，显著减少对专用结构的依赖并刷新多项记录；更关键的是，在BSS/SD等生成任务中，我们首次以重建目标微调并证实预训练权重依然有效，拓宽了范式边界。
%
\begin{itemize}
\item We present EMind, a purpose-built foundation model for electromagnetic signals that bridges the gap between large-scale pre-training paradigms and the EM signal modality, establishing a solid basis for superior representation learning.
\item We design an architecture tailored to electromagnetic data, introducing a signal discretization mechanism, variable-length input handling, and efficient training strategies for massive heterogeneous datasets, all addressing the unique characteristics of variable sequence length EM signals and the coexistence of temporal sparsity and continuity.
\item We construct the largest-to-date EM signal dataset, spanning diverse signal classes, rich attribute labels, and multiple downstream tasks, delivered with unified normalization and high-efficiency storage to create a robust data foundation for large-scale pre-training and universal feature learning.
\item EMind systematically demonstrates the broad applicability of pre-train–fine-tune to EM signal tasks, markedly reducing reliance on task-specific designs and setting new SOTA results. Critically, we are the first to adopt reconstruction-oriented fine-tuning for generative tasks such as BSS and SD, empirically showing that pre-trained weights retain their efficacy and thereby extending the paradigm’s frontier.
\end{itemize}

\section{Related Works}
\label{related}

% 多数都用图，图的好处vit，提炼信息，丢失信息(图没有相位信息)
% raw的好处：保留最多原始信息，模型越强，数据越原始就可以最大化利用信息，上升空间越大
% %通用领域-专用领域
%mim-自监督预训练方法，mae
% 医疗，遥感领域有，电磁尚且空白：% radiollm

This section provides a review of existing literature relevant to Foundation Models in communication and sensing, categorized into three key areas: training datasets and method design.

\subsection{Training Datasets for EM Foundation Models}
% 电磁基础模型在很大程度上依赖于大规模、多样化且高质量的数据集来进行预训练 \cite{zhou2024comprehensive, fontaine2024towards， sheng2025wireless， liang2024foundation}。具体而言，电磁基础模型侧重于利用广泛的无线数据集进行预训练，以获取与任务无关的表示。然而，在无线通信环境中收集此类数据集面临着独特的挑战，这是由于射频（RF）环境的复杂性和动态性，其中包括信号条件、干扰模式和传播特性方面的变化 \cite{zhang2019deep， guler2025multi}。
EM Foundation Models heavily depends on the availability of large-scale, diverse, and high-quality datasets for pre-training \cite{zhou2024comprehensive, fontaine2024towards, sheng2025wireless, liang2024foundation}. Specifically, the EM foundation models focus on pre-training with extensive wireless datasets to acquire task-agnostic representations. However, collecting such datasets in wireless communication contexts presents unique challenges due to the complex and dynamic nature of radio frequency (RF) environments, which include variations in signal conditions, interference patterns, and propagation characteristics \cite{zhang2019deep, guler2025multi}.

% 为了构建可靠的数据集，大多数电磁基础模型的开发在历史上都依赖于合成或模拟数据。例如，WirelessGPT \cite{yang2025wirelessgpt} 是基于大规模无线信道数据集进行预训练的，该数据集包括了公开可用的数据集以及自开发的数据集，如 Traciverse、SionnaRT \cite{hoydis2022sionna} 和 DeepMIMO \cite{alkhateeb2019deepmimo}。同样，WiMAE \cite{guler2025multi} 利用 DeepMIMO 数据集生成了超过一百万个信道样本，用于在各种场景下进行预训练。虽然模拟数据集提供了控制和可扩展性，但它们可能无法完全捕捉现实世界电磁环境的复杂性和不可预测的变异性。
To construct reliable dataset, most EM foundation models development has historically leveraged synthetic or simulated data. For instance, WirelessGPT \cite{yang2025wirelessgpt} was pre-trained on a large-scale wireless channel dataset, encompassing both publicly available and self-developed datasets like Traciverse, SionnaRT \cite{hoydis2022sionna}, and DeepMIMO \cite{alkhateeb2019deepmimo}. Similarly, WiMAE \cite{guler2025multi} utilized the DeepMIMO dataset to generate over a million channel samples for pre-training across various scenarios. While simulated datasets offer control and scalability, they may not fully capture the intricacies and unpredictable variabilities of real-world EM environments.

% 为了提高预训练数据集的质量，一些研究采用了真实世界的数据。例如，WavesFM \cite{aboulfotouh20256g} 仅基于真实世界的数据进行了预训练，这些数据包括射频（RF）光谱图、WiFi 信道状态信息（CSI）样本以及从不同地理位置和场景收集的第五代（5G）CSI 样本。此外，SpectrumFM \cite{zhou2025spectrumfm} 是一种用于智能频谱管理的基础模型，它从各种来源汇集了一个全面的数据集，包括公开可用的开源数据集，如 RML2018.01A \cite{o2018over} 和 TechRec \cite{fontaine2019towards}，以及在实际应用场景中收集的信号。这些努力突显了向更真实和多样化的数据来源转变的趋势，以构建强大的且普遍适用的电磁基础模型。
To enhance the quality of pre-training datasets, some studies incorporate real-world data. For example, WavesFM \cite{aboulfotouh20256g} was pre-trained solely on real-world data, including Radio Frequency (RF) spectrograms, WiFi Channel State Information (CSI) samples, and 5th Generation (5G) CSI samples collected from diverse geographical locations and scenarios. Additionally, SpectrumFM \cite{zhou2025spectrumfm}, a foundation model for intelligent spectrum management, compiled a comprehensive dataset from various sources, including publicly available open-source datasets like RML2018.01A \cite{o2018over} and TechRec \cite{fontaine2019towards}, as well as signals collected in real-world practice scenarios. These efforts highlight a shift towards more realistic and diverse data sources to build robust and universally applicable EM Foundation Models.

% 基础模型的性能会受到其预训练数据集的影响很大。然而，先前研究中所使用的数据集在多样性及规模方面往往存在局限性，从而限制了这些模型在下游任务中的泛化能力。
The performance of foundation models is significantly impacted by their pre-training datasets. However, datasets utilized in prior work often exhibit limitations in terms of diversity and volume, consequently constraining the models' generalization capabilities on downstream tasks.

\subsection{Design of Foundation Models towards Electromagnetic Signals}

% EM 基础模型的架构设计对于有效捕捉无线数据中固有的复杂时空和频率相关性至关重要，并且能够实现高效的多任务学习 \cite{awais2025foundation，han2024foundation,hong2023spectralgpt} 。
The architectural design of EM Foundation Models is crucial for effectively capturing the complex spatio-temporal and frequency correlations inherent in wireless data, and for enabling efficient multi-task learning \cite{awais2025foundation,han2024foundation,hong2023spectralgpt}.

% 受自然语言处理（NLP）和计算机视觉（CV）中 Transformer 架构成功应用的启发，许多 EM 基础模型采用了类似的原理 \cite{vaswani2017attention,yang2025wirelessgpt,guler2025multi}。例如，无线 GPT \cite{yang2025wirelessgpt} 采用了一种基于 Transformer 的基础模型来捕捉无线信道数据中的空间、时间以及频率相关性。同样地，WavesFM \cite{aboulfotouh20256g} 利用了一种基于视觉 Transformer（ViT）的架构，通过自监督学习来学习丰富的无线信号表示，特别是通过一种掩码无线建模（MWM）方法。此外，一些研究专注于改进信号数据的架构设计。例如，SpectrumFM \cite{zhou2025spectrumfm} 引入了一种新颖的混合编码器架构，该架构将卷积神经网络（CNN）用于局部特征提取与多头自注意力（MHSA）机制协同结合。
Inspired by the success of transformer architectures in Natural Language Processing (NLP) and Computer Vision (CV), many EM foundation models adopt similar principles \cite{vaswani2017attention,yang2025wirelessgpt,guler2025multi}. For instance, WirelessGPT \cite{yang2025wirelessgpt} employed a Transformer-based foundation model to capture spatial, temporal, and frequency correlations in wireless channel data. Similarly, WavesFM \cite{aboulfotouh20256g} utilized a Vision Transformer (ViT)-based architecture that learns rich wireless signal representations via self-supervised learning, particularly through a Masked Wireless Modeling (MWM) approach. Furthermore, some research focuses on enhancing architectural design for signal data. For instance, SpectrumFM \cite{zhou2025spectrumfm} introduced a novel hybrid encoder architecture that synergistically combines Convolutional Neural Networks (CNNs) for localized feature extraction with Multi-Head Self-Attention (MHSA) mechanisms.

% 除了结构设计之外，一些工作还探索了多种预训练方法。例如，WiMAE \cite{guler2025multi} 采用了掩码信号重建来进行预训练。此外，ContraWiMAE \cite{guler2025multi} 通过引入对比学习目标来扩展这一方法，以增强判别性特征学习。除了基于掩码的预训练之外，还有一些方法利用了混合预训练策略。例如，SpectrumFM \cite{zhou2025spectrumfm} 是通过两个自监督学习任务进行预训练的：掩码重建和下一个槽位信号预测。这种方法使模型能够学习潜在结构和内在相关性，从而提高其特征提取能力和鲁棒性。
Beyond architectural design, some works explore diverse pre-training methodologies. For instance, WiMAE \cite{guler2025multi} adopted masked signal reconstruction for pre-training. Furthermore, ContraWiMAE \cite{guler2025multi} extended this approach by incorporating a contrastive learning objective to enhance discriminative feature learning. In addition to masking-based pre-training, some methods utilize hybrid pre-training strategies. For example, SpectrumFM \cite{zhou2025spectrumfm} was pre-trained via two self-supervised learning tasks: masked reconstruction and next-slot signal prediction. This approach enables the model to learn latent structures and intrinsic correlations, thereby improving its feature extraction capabilities and robustness.

% 然而，不同数据集之间存在的差异性数据格式和规模限制了上述研究中数据的利用效率。
However, the disparate data formats and scales across various datasets limit the data utilization efficiency of the aforementioned studies.

\section{Problem Formulation and Challenges}
\begin{figure}[H]
  \centering
  \begin{minipage}{\linewidth}
    \centering
    %---- 子图 A ----
    \includegraphics[width=\linewidth]{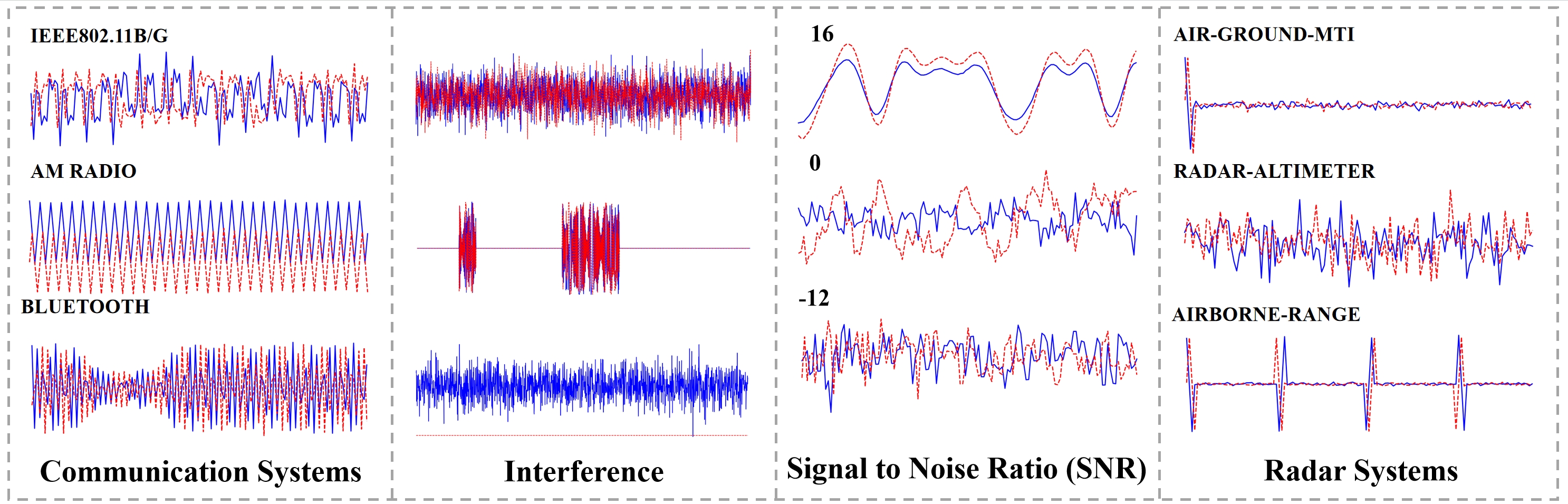}
    \subcaption{Diverse signal types with intricate semantics exhibit a wide spectrum of characteristics.}
    \label{fig:chall1}
    %---- 子图 B ----
    \includegraphics[width=\linewidth]{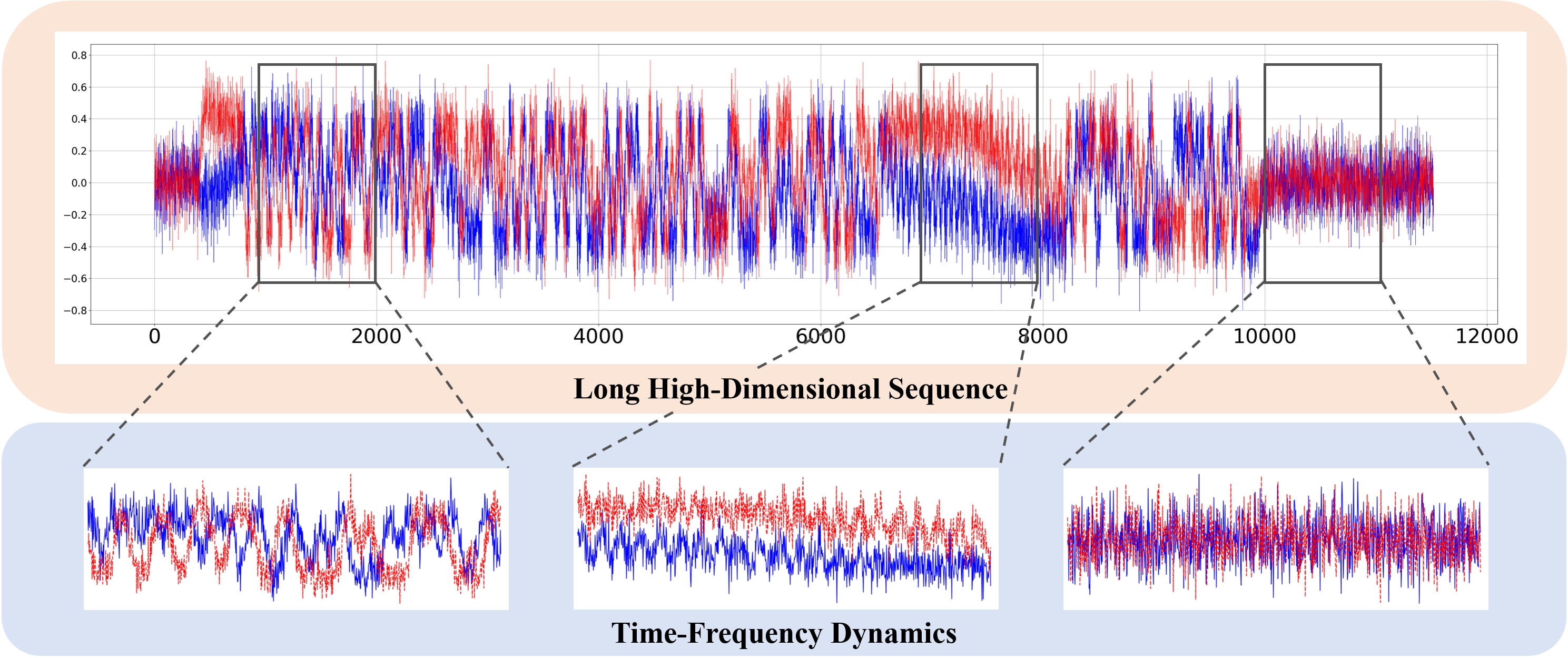}
    \subcaption{Long high-dimensional signals show time-frequency dynamics.}
    \label{fig:chall2}
  \end{minipage}
  \caption{Visualization of signals exhibits their heterogeneous and complicity.}
  \label{fig:challenges}
\end{figure}
%
%电磁信号是一类特殊的时间序列数据，主要由通信台站、雷达辐射源等电磁目标发射，经无线空间传播到接收端。在接收链路中，这些信号首先经过射频（RF）接收模块进行下变频、滤波等模拟处理，转换为基带信号，再以复数形式表示为 IQ（In-phase \& Quadrature）采样数据。IQ 信号是电磁信号在接收端数字域的标准表示方式，保留了原始信号的幅度与相位特征，适用于特征提取、分类、识别等任务（新领域数据特征：数据怎么来的，iq数据特征-iq正交，数据是以什么形式出现）
Electromagnetic signals constitute a specialized class of time-series data, typically emitted by electromagnetic sources such as communication stations and radar systems. After propagating through the wireless medium, these signals are received and processed by the receiver front-end, where they undergo analog operations including downconversion and low-pass filtering to obtain the baseband signal. The resulting baseband signal is then represented in complex form as in-phase and quadrature (IQ) samples. This IQ representation, which preserves both amplitude and phase information of the original waveform, serves as the canonical digital form of EM signals at the receiver side. Specifically, a real-valued EM signal $x(t)$ with a carrier frequency $f_c$ can be expressed as:
\begin{equation}
x(t)=I(t) \cos \left(2 \pi f_c t\right)-Q(t) \sin \left(2 \pi f_c t\right),
\end{equation}
\noindent where $I(t)$ and $Q(t)$ represent the in-phase and quadrature-phase components of the signal, respectively. The baseband signal can be further discretized with a sampling  rate $f_s$, resulting in a discrete IQ sequence, which is,
\begin{equation}
s[n]=I[n]+j Q[n], \quad n=0,1,2, \ldots, N-1.
\end{equation}
\noindent Here, $n$ represents the length of the sequence. 
%随着电磁环境复杂性的不断增加，电磁信号的种类和数量迅速增长。由于各种任务的特征和目标存在显著差异，导致单任务学习模型在跨任务泛化时表现不佳，训练效率较低。因此，迫切需要学习一种能够捕获通用特征的判别表示，以增强下游任务的泛化能力，而开发这样的电磁信号基础模型仍面临以下挑战：
% 问题：目标是学习到通用的特征表示，有xxxxxxx任务（公式化），通过这个表征可以获得好的下游任务的结果。
As the complexity of the electromagnetic environment continues to increase, the types and quantity of electromagnetic signals are rapidly growing. Due to the significant differences in the characteristics and objectives of various tasks, single-task learning models perform poorly in cross-task generalization and have low training efficiency. Therefore, there is an urgent need to learn discriminative representations that capture general features to enhance the generalization ability of downstream tasks. However, developing such a foundational model for EM signals still faces the following challenges:

% Challenge 1: How to 构建统一的预训练数据集
% Challenge 2: How to 构建适应电磁信号的大model
\textbf{Challenge 1: Constructing a unified pre-training dataset for electromagnetic signals.}
% 电磁信号种类繁多，且其调制方式和传播特性具有显著的异质性，这使得电磁信号的语义结构异常复杂，既包含短时瞬态特征，又具有长程依赖关系。同时，在高带宽或长时窗的观测条件下，电磁信号容易产生极长的高维序列（如图{fig:challenges}所示）。因此，构建一个涵盖多种信号类型的预训练数据集面临着数据筛选、对齐和统一的挑战，必须确保这些信号具有共同的基础，才能使得有效的训练成为可能。
EM signals are diverse, and their modulation methods and propagation characteristics exhibit significant heterogeneity, which makes the semantic structure of EM signals exceptionally complex, containing both short-term transient features and long-range dependencies. Additionally, under high bandwidth or long time-window observation conditions, EM signals tend to generate extremely long high-dimensional sequences (as shown in Figure~\ref{fig:challenges}). Therefore, constructing a pretraining dataset that covers multiple signal types faces challenges in data selection, alignment, and standardization. It is essential to ensure that these signals share a common foundation, so that effective training becomes possible.

\begin{figure*}[hbp]
  \begin{flushright}
    \begin{minipage}{1.\linewidth}
      \flushright
      \includegraphics[width=\linewidth]{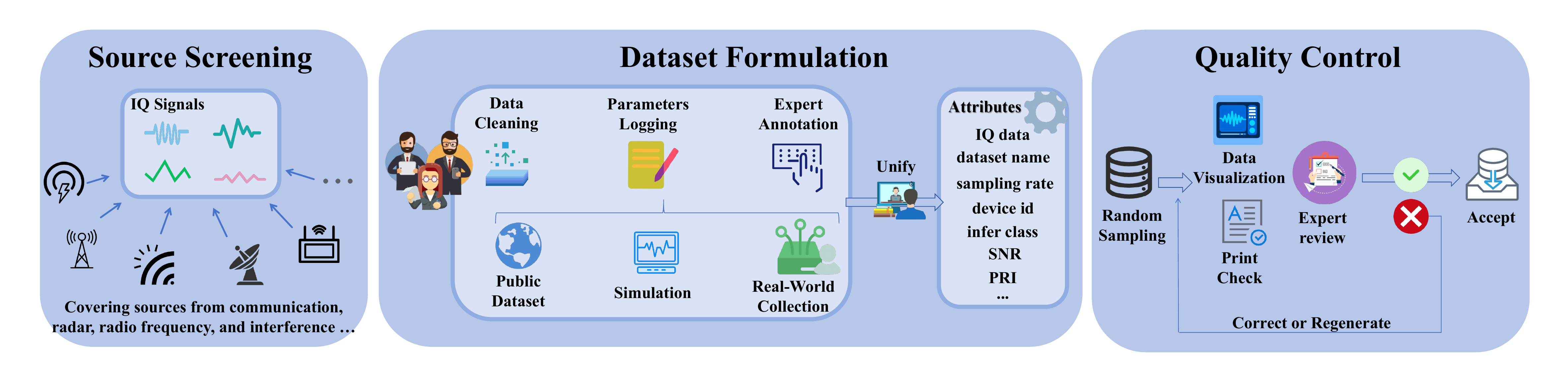}
      \caption{Pipeline of EM pre-train dataset cosntruction. Our pipeline comprises 3 stages all led by experts: Source Screening, Dataset Formulation, and Quality Control.}
      \label{fig:pipe}
    \end{minipage}
  \end{flushright}
\end{figure*}

\textbf{Challenge 2: Developing an effective model architecture toward electromagnetic signals.}
% 电磁信号作为一种独特的模态，不同类型的信号之间存在显著的语义差异，各个信号阶段通常承载着与特定任务相关的语义信息。因此，设计一个既能够兼顾电磁信号的物理属性，又能有效表达其语义信息的网络架构，并高效地应用预训练-微调策略，是构建适用于电磁信号的基础模型所面临的重大挑战。
EM signals, as a unique modality, exhibit significant semantic differences between different types of signals. Each signal stage typically carries semantic information related to specific tasks. Therefore, designing a network architecture that can both account for the physical properties of EM signals and effectively express their semantic information, while efficiently applying the pretraining-finetuning strategy, presents a major challenge in constructing a foundation model suitable for EM signals.

\section{Dataset} % 电磁信号大规模数据集

% 数据集构建流程如下：我们首先系统整合了来自通信、雷达、射频和干扰等领域的多源数据集，所有信号均以原始IQ（同相-正交）格式保存。为确保数据质量并实现统一的大规模预训练，我们对原始数据进行了严格清洗，由领域专家完成详尽的参数记录和标注工作，并将数据统一转换为n×2维格式存储为Parquet文件。此外，我们还通过随机抽样的可视化审核方式对数据进行质量验证。经过这一系列严谨的处理流程，最终建成了当前已知规模最大的面向机器学习应用的电磁信号数据集，如图{fig:pipe}所示。（被动语态一般现在时）
The dataset construction pipeline is as follows: multi-source datasets from communications, radar, radio frequency, and interference domains are systematically integrated, with all signals saved in their raw IQ (in-phase and quadrature) format. To ensure data quality and enable unified large-scale pretraining, data cleansing is performed on the raw data; detailed parameter recording and labeling are completed by domain experts, and the data are uniformly converted into an n×2 dimensional format and stored as Parquet files. In addition, quality validation is conducted through randomized visual inspections. Through this series of stringent procedures, the largest known EM signal dataset for machine learning applications is ultimately constructed, as show in Figure~\ref{fig:pipe}.
\begin{table*}[t]
  \caption{EM Pre-train Dataset. Bold indicates self-collected data, the others are publicly available datasets.}
  \label{tab:pretrain}
  \centering
  \resizebox{1.0\textwidth}{!}{
  % \begin{threeparttable}
  \renewcommand{\arraystretch}{1.0}  % 增大表格行间距
  \fontsize{10}{10}\selectfont
  \begin{tabular}{lrrrrrrrr}
    \toprule
     & & & Signal & Signal & Sampling  & Data* & Number of \\
    Dataset Name  & Transmitter & Receiver &  Type & Length  & Rate & Source & Samples \\
    \midrule
    \bf EM-Comm & \bf  a USRP X310 &  \bf - & \bf  Category 14 modulated signals,  & \bf  1,024 & \bf  20 MHz &  \bf  Real-world & \bf 2,747,000 \\
     &  &  & \bf SNR: -20$\sim$18db &  &  &  &  \\
    HisarMod2019.1 \cite{tekbiyik2019hisarmod} & -  & - &  Category 26 modulated signals, & 1,024  & 1 MHz & Simulation &  780,000   \\
     &  &  & SNR: -20$\sim$18db  &  &  &  &  \\
    Panoradio HF \cite{scholl2019classification} & - & - & Category 18 modulated signals, & 2,048 & 6 kHz & Simulation &  172,800   \\
     &  &  & SNR: -10$\sim$25d  &  &  &  &  \\
    RadarCommDataset \cite{jagannath2021dataset} & a USRP N210 & USRP N210 & Category 6 modulated signals, & 128 & 10 MHz & Real-world & 860,361  \\
     &  &  & Category 8 signal types, &  &  &  \&Simulation &  \\
     &  &  & SNR: -20$\sim$18db &  &  &  &  \\
    \bf EM-RadarParaIQSim & \bf  - & \bf  - & \bf  Category 2 radar types, & \bf  320/384 & \bf  10/12 MHz & \bf Simulation  & \bf  300,000  \\
    &  &  & \bf Category 3 parameters, & \bf  400/480 &  &  &  &  \\
    &  &  & \bf SNR: 6$\sim$12db &  &  &  &  \\
    \bf EM-Radar & \bf - & \bf  - & \bf  10 radar emitters, & \bf 1,024 & \bf  5 MHz  & \bf  Simulation & \bf  327,680 \\
    &  &  & \bf including 5 radar waveforms, &  &  &  &  \\
    &  &  & \bf and 3 radar system parameters &  &  &  &  \\
    &  &  & \bf SNR: -10$\sim$20db &  &  &  &  \\
    WiSig \cite{hanna2022wisig} & 174 WiFi Emitter &  41 USRP B210/N210/X310 & IEEE802.11a/g & 256  & 25 MHz  & Real-world &  18,243,630  \\
    Northeastern RF \cite{al2020exposing} & 20 USRP devices & - & IEEE802.11a/g & 288 & 20 MHz  & Real-world &  17,930,880  \\
    POWDER \cite{reus2020trust} & 4 USRP X310 & USRP B210 &  5G, WiFi, LTE & 512 & 5/7.68 MHz  & Real-world &  1,044,472 \\ 
    Transmitter Classification \cite{morin2019transmitter} & 21 USRP devices & USRP N2932 & IEEE 802.15.4 & 600  & 5 MHz  & Real-world &  11,928,835 \\
    LoRa RF Datasets \cite{elmaghbub2021lora} & 25 Pycom devices & USRP B210 & LoRa  & 4,096 & 1 MHz & Real-world  &  13,498,730 \\
    Mono Receiver \cite{liu2020zero} & Over 140 civil aircrafts & USRP B210 & ADS-B & 974 & 8 MHz  & Real-world &  30,367  \\
    DroneRFa \cite{2024dronerfa} & 24 categories of UAVs & USRP & Communication RF & 1,024 & 100 MHz & Real-world &  13,132,770 \\
    \bf EM-Infer-Radar-v2 & \bf - & \bf - & \bf  Category 12 radar interference types & \bf  2,048 & \bf 20 MHz & \bf Simulation &  \bf 120,000  \\
    &  &  & \bf  ISR: 30$\sim$60db &  &  &  &  \\
    \midrule
    TOTAL &&&&& &&  \bf 81,117,525 \\
    \bottomrule
  \end{tabular}
  % \begin{tablenotes}
  %     \item[*] R refers to real-world data, while S refers to simulated data.
  % \end{tablenotes}
  % \end{threeparttable}
  }
\end{table*}

\subsection{Source Screening}

% 我们构建的数据集涵盖了多种信号类型，其核心原则是最大化信号与任务的多样性，以确保训练数据的广泛性和有效性。为了实现这一目标，我们系统地组织了多个公开可用的信号数据集，并通过使用自行收集的数据补充公共数据来解决类别缺失或样本不足。总的来说，如表~\ref{tab:pretrain}所示，我们的预训练数据集由14个数据集组成，覆盖了广泛各异的采集设备、信号类型、信号长度、采样率等，其中，10个是公开数据库，样本量为77,622,845个，其余4个是自采数据集（表中粗体，已EM-为前缀），总样本量为3,494,680。需要强调的是，针对公开数据集，我们仍然进行了认真的选择和预处理。最终，我们构建的预训练数据集的总规模达到81,117,525个样本，使其成为迄今为止已知的最大的电磁信号预训练数据集，其详细的数据集配置参见\hl{appendix}。
The dataset we constructed comprises a wide variety of signal types, with the core principle being the maximization of signal and task diversity to ensure both the breadth and effectiveness of the training data. To achieve this, we systematically curated multiple publicly available signal datasets and supplemented the public data with self-collected data to address missing categories or insufficient samples. As shown in Table~\ref{tab:pretrain}, our pretraining dataset consists of 14 datasets, covering a wide range of acquisition devices, signal types, signal lengths, sampling rates, and other factors. Of these, 10 are publicly available databases, totaling 77,622,845 samples, 
while the remaining four datasets are self-collected (bolded in the table, prefixed with 'EM-'), totaling 3,494,680 samples. It is worthy to note that, even for the publicly available datasets, we conducted thorough selection and pre-processing. Ultimately, the total volumn of our pretraining dataset reaches 81,117,525 samples, making it the largest known EM signal pretraining dataset to date. Detailed dataset configurations can be found in the Appendix.

\subsection{Expert Annotations}
% 对于电磁信号的标注工作，公开数据集的真值标签通常由原始发布方提供。对于仿真数据集，标签在数据生成过程中自动产生，来源于预设的调制方式、信道参数或雷达波形配置。对于自主采集的数据集，标签由采集流程控制，通常根据实验人员记录的设备类型、采样距离、频段、调制方式、信号强度等信息生成。
For the annotation of EM signals, the ground truth labels for publicly available datasets are typically provided by the original data publishers. In the case of simulation datasets, the labels are automatically generated during the data generation process, based on predefined modulation schemes, channel parameters, or radar waveform configurations. For self-collected datasets, the labels are controlled by the data acquisition process, usually generated based on information recorded by the experimenters, such as device types, sampling distances, frequency bands, modulation types, and signal strength.
% 基于这些标注标准，我们对数据集进行了统一整理，并提取了以下17个关键属性：\verb|iq_data|、\verb|dataset_name|、\verb|sampling_rate|、\verb|device_id|、\verb|transmission_id|、\verb|infer_class|、\verb|snr|、\verb|isr|、\verb|modulation_type|、\verb|radar_waveform_type|、\verb|pri|、\verb|pulse_time_delay|、\verb|number_of_pulses|、\verb|pulse_width|、\verb|bandwidth|、\verb|amplitude|、\verb|radar_segmentation_type| 等。如果某一数据集中缺失某个属性，则该属性被填为 \verb|None|.
Based on these principles, we have standardized 17 key attributes: iq data, dataset name, sampling rate, device id, transmission id, infer class, signal-to-noise ratio (SNR), interference-to-signal ratio (ISR), modulation type, radar waveform type, pulse repetition interval (PRI), pulse time delay, number of pulses, pulse width, band width, amplitude, and radar segmentation type. If a specific dataset lacks one or more of these fields, the corresponding entries are assigned a default value of \verb|None|. 

\subsection{Quality Control}
% 为了确保数据的完整性与任务相关性，建立了可靠的质量控制流程。首先，从数据库中随机抽取 IQ 数据，打印其全部参数并绘制波形，结合采样率、幅值范围等对数据格式和内容进行核查，同时排查是否存在异常填充等问题。对于发现的任何异常或差异，均予以标记，并提交领域专家复核，查明原因后对错误数据进行修正或重新生成，确保数据的准确性。该闭环质量控制机制实现了数据集的持续优化，有效保障了整体数据的可靠性。
To ensure the integrity of the data and its relevance to the task, a reliable quality control process has been established.
First, IQ data is randomly sampled from the dataset, with all parameters printed and waveforms plotted. This process involves checking the data format and content, considering factors such as sampling rate and amplitude range, while also inspecting for issues such as anomalous padding.
Any anomalies or discrepancies discovered are marked and submitted for expert review. After identifying the cause, erroneous data is either corrected or regenerated to ensure accuracy. This closed-loop quality control mechanism enables the continuous optimization of the dataset, effectively ensuring the overall reliability of the data.

\subsection{Analysis}
\begin{table*}[h!]
  \caption{Comparison of pre-train datasets}
  \label{tab:comp_pre_train}
  \centering
  \resizebox{0.9\textwidth}{!}{
  \begin{threeparttable}
  \begin{tabular}{lccrrr}
    \toprule
    Pre-train & Dataset Source & Include   & Signal  & Scene  & Number of \\
    Dataset & Volumn & Self-constructed  &  Length &  Type & Samples   \\
    \midrule
    RF-Diffusion \cite{chi2024rf}  & 2 & $\checkmark$ & 512 & radio frequency & 33 k\\
    \midrule
     &  && &  communication  &   \\
    ACSF-TMAE \cite{chen2024generative} & 4 & $\times$ & 128 \& 3,000 & \& radio frequency  &  1 million \\
    \midrule
    RadioLLM \cite{chen2025radiollm} &  4 & $\times$ & 128 \& 1,024 & communication  &  4 million\\
    \midrule
     &  && &  communication  & $8$-$10$ million$^* + $ \\
    SpectrumFM \cite{zhou2025spectrumfm} & 3 & $\checkmark$ & 128 & \& radio frequency & 25 GB$^{\dagger}$ \\
    \midrule
     & & && communication \& radar &  \\
     &  &&& \& radio frequency  &  \\
    EMind & 14  & \checkmark & 128$\sim$4096 & \& interference & 81 million \\
    \bottomrule
  \end{tabular}
  \begin{tablenotes}
      \item[*] In \cite{zhou2025spectrumfm}, the pre-training data volume for TechRec is estimated according to \url{https://github.com/JaronFontaine/Technology-Recognition-dataset-of-real-life-LTE-WiFi-and-DVB-T?tab=readme-ov-file}
      \item[$\dagger$] 25GB is a self-collected dataset without length unspecified.
  \end{tablenotes}
  \end{threeparttable}
  }
\end{table*}
\begin{table}[h!]
  \caption{Main statistics of our dataset}
  \label{tab:statistics}
  \centering
  \resizebox{0.5\textwidth}{!}{
  % \begin{threeparttable}
  \begin{tabular}{lr}
    \toprule
    Statistic & Number of Samples   \\
    \midrule
    Task type & \\
    - Communication & 3,699,800 \\
    - Radar & 1,488,041\\
    - Radio frequency & 75,809,684\\
    - Interference & 120,000\\
    \midrule
    Attributes & \\
    - Modulation &  4,560,161 \\
    - Radar Waveform & 627,680 \\
    - Signal-to-Noise Ratio (SNR) & 5,218,208\\
    - Interference class & 120,000 \\
    - Interference-to-Signal Ratio (ISR)  & 120,000 \\
    - Band width & 327,680\\
    - Pulse repetition interval (PRI)  & 627,680 \\
    - Pulse width & 627,680 \\
    - Device id & 75,809,684\\
    - Transmission id & 17,930,880\\
    \midrule
    Signal Length & \\
     - 128 & 860,361 \\
     - 256 & 18,243,630 \\
     - 288 & 17,930,880 \\
     - 320 & 75,000 \\
     - 384 & 75,000 \\
     - 400 & 75,000 \\
     - 480 & 75,000 \\
     - 512 & 1,044,472\\
     - 600 & 11,928,835\\
     - 974 & 30,367\\
     - 1,024 & 16,987,450\\
     - 2,048 & 292,800\\
     - 3,840 & 12,399\\
     - 4,096 & 13,498,730\\
    \bottomrule
  \end{tabular}
  % \begin{tablenotes}
  %     \item[*] The pre-training data volume for TechRec is estimated based on \url{https://github.com/JaronFontaine/Technology-Recognition-dataset-of-real-life-LTE-WiFi-and-DVB-T?tab=readme-ov-file}
  % \end{tablenotes}
  % \end{threeparttable}
  }
\end{table}
%
% 我们的预训练数据库在场景类型、信号长度及数据规模等方面具有显著优势，涵盖了4个主要场景、8种信号类型、29种调制方式、9种雷达波形和12种干扰类型。此外，数据库能够处理的信号长度范围广泛，采样点从128到4096均可覆盖，其与其他预训练数据库的对比如表所示。
Our pre-training dataset exhibits pronounced advantages in terms of scene diversity, signal length, and data volume. It encompasses four principal operational scenarios (communication, radar, radio frequency and interference), 
eight distinct signal modalities (Airborne detection radar, Airborne range radar, Air-Ground MTI radar, Ground mapping radar, Radar Altimeter, SATCOM, AM Radio, Short-range wireless),
twenty-nine modulation formats (BPSK, QPSK, 8PSK, 16PSK, 32PSK, 64PSK, 4QAM, 8QAM, 16QAM, 32QAM, 64QAM, 128QAM, 256QAM, 2FSK, 4FSK, 8FSK, 16FSK, 4PAM, 8PAM, 16PAM, AM-DSB, AM-DSB-SC, AM-USB, AM-LSB, FM, PM, morse, psk31, psk63), 
nine radar waveforms Coherent Pulse Train, Barker Code, Polyphase Barker Code, Frank Code, Linear Frequency Modulated (LFM), Rectangular, Phase Coded, Stepped FM and Custom FM), 
and twelve interference classes (Pure Noise, Intermittent Sampling Forwarding Interference, Spot-Jamming Interference, Blocking Interference, Frequency-Sweeping Interference, Range-Fooling Interference, Dense False Target Interference, Smart Noise Interference, Chaff Interference, Chaff Interference Combined with Intermittent Sampling Forwarding Interference, Dense False Target Interference Combined with Smart Noise Interference, and Range-Fooling Interference Combined with Frequency Modulated Frequency-Sweeping Interference).
Furthermore, the dataset accommodates signal lengths spanning 128 to 4,096 sampling points. A comparative summary against other available pre-training dataset is provided in Table~\ref{tab:comp_pre_train} and Figure~\ref{fig:comp}.

\subsubsection{Diversity of task types} 
% 我们的数据集包含了多种任务类型，包括通信数据集EM-Comm、HisarMod2019.1和Panoradio HF；雷达数据集RadarCommDataset、EM-RadarParaIQSim和EM-Radar；干扰与异常数据集EM-Infer-Radar-v2；无线射频数据集WiSig Raw、Northeastern RF、POWDER、Transmitter Classification、LoRa RF Datasets、LoRa RFFI、Mono Receiver和DroneRFa，涵盖了从WiFi、LoRa到ADS-B、UAV等多种协议与设备。
Our dataset includes a variety of task types, including communication datasets EM-Comm, HisarMod2019.1, and Panoradio HF; radar datasets RadarCommDataset, EM-RadarParaIQSim, and EM-Radar; interference dataset EM-Infer-Radar-v2; and wireless RF datasets WiSig Raw, Northeastern RF, POWDER, Transmitter Classification, LoRa RF Datasets, Mono Receiver, and DroneRFa, covering a wide range of protocols and devices from WiFi, LoRa to ADS-B, UAV, and more.

\begin{figure*}[htbp]
  \begin{flushright}
    \begin{minipage}{0.95\linewidth}
      \flushright
      \includegraphics[width=\linewidth]{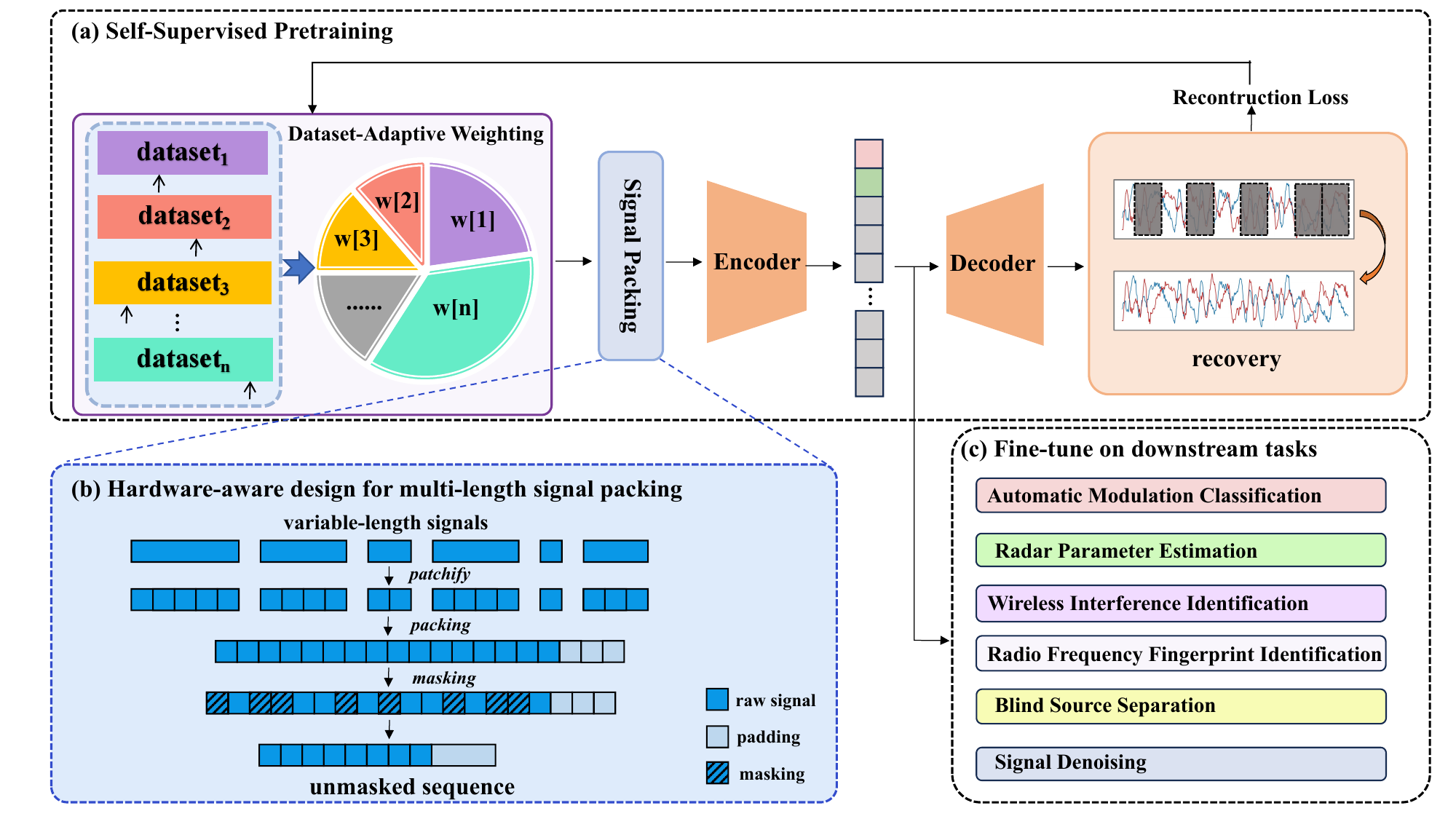}
      \caption{EMind (a) Large-scale multi-scenario self-supervised pre-training with dataset-adaptive weighting. (b) Low-redundancy length adaptive multi-signal packing and per-sample masking. (c) Fine-tune on multiple downstream tasks with support for arbitrary-length signal inference.}
      \label{fig:model}
    \end{minipage}
  \end{flushright}
\end{figure*}

\subsubsection{Comprehensive annotations} 
% 我们对数据集提供了详细的标注信息，包括通信信号的调制类型；雷达信号的波形类型、脉宽、脉冲重复周期等；射频信号的辐射源标识，带宽，以及干扰信号的干扰类型，干信比等特征。所有数据均标注了其采样率信息（通常从千赫兹（kHz）至数百兆赫兹（MHz）不等），这一属性对于电磁信号的处理至关重要。此外，所有可获取的数据都标注了信噪比（SNR）信息，这些详细的标注确保了数据的高质量和全面性。
We provide detailed annotation information for the dataset, including the modulation type of communication signals; waveform type, pulse width and pulse repetition period for radar signals; device id and bandwidth for RF signals; and interference type and interference-to-signal ratio (ISR) for interference signals.
All data are annotated with sampling rate, which typically ranges from kilohertz (kHz) to several hundred megahertz (MHz), a critical attribute for processing EM signals. Additionally, all available data are annotated with Signal-to-Noise Ratio (SNR). These comprehensive annotations ensure the high quality and completeness of the dataset.

\subsubsection{Length of Samples} 
%
% 我们的预训练数据集包含的样本长度从128到4096不等，这一广泛的长度范围使得模型能够在更大范围的时序数据上进行训练，相比之下，现有的预训练数据库通常仅包含单一场景数据集或几个固定长度的数据集，如表{tab:statistics}所示。这种灵活的序列长度设计为可变长信号的推理提供了可能性，允许模型更好地适应不同长度的复杂信号。特别是在电磁信号等实际应用中，信号的时序长度和频谱特征往往具有高度的变动性，通过支持可变长序列，我们的预训练数据集能够提高模型在面对真实世界数据时的泛化能力，使其更有效地捕捉不同信号的特征和依赖关系，其统计表如表x所示。
Our pre-training dataset covers signal lengths ranging from 128 to 4,096 samples, which enables the model to train on a wider range of dataset. As detailed in Table \ref{tab:comp_pre_train}, existing datasets are typically restricted to a single operational scenario or composed of fixed-length signals. Our flexible signal-length design enables the potential for arbitrary length inference, allowing the model to better adapt to complex signals of varying durations. Such variability markedly improves generalization to real-world data, allowing the network to robustly capture both transient features and long-range dependencies across diverse EM signals; detailed statistics are provided in Table \ref{tab:statistics}.

\section{EMind: ElectroMagnetic Signal Foundation Model}\label{meth}
% 针对challenge 1：直接表明我们的基础模型是如何解决challenge 1的？一大段表述。
% 为应对“电磁信号语义复杂、异构性强”这一核心挑战，本文在方法层面提出并贯彻“大规模预训练 + 轻量微调”的统一范式，如图\ref{fig:model}所示。具体而言，首先基于 Transformer 编码器-解码器架构，采用掩码自编码（MAE）在大规模无标签多场景 IQ 流上进行预训练；通过高达 75 % 的掩码率迫使模型在极端信息缺失条件下重建iq信号，从而习得对调制类型、传播信道均不敏感的通用连续-稀疏表征。随后，在下游任务微调阶段仅训练轻量级任务头。该两阶段策略将复杂瞬态语义、稳态特征及异构调制差异统一映射到共同高维流形，实现少标签即可快速适配、跨域仍具强泛化能力的多任务电磁信号高质量建模。
To address the fundamental challenge of high heterogeneity and semantic complexity inherent in EM signals, we propose to implement a unified pre-training and fine-tuning paradigm, as shown in Fig.~\ref{fig:model} (a). Concretely, we first adopt a Transformer encoder–decoder architecture and perform Masked Auto-Encoding (MAE) on massive unlabeled multi-scenario IQ streams. By imposing an aggressive masking ratio of 75 \%, the model is forced to reconstruct the IQ signal under extreme information scarcity, thereby learning modulation- and channel-agnostic universal continuous–sparse representations. In the subsequent downstream fine-tuning stage, a task-specific head is fine-tuned. This two-stage strategy projects intricate transient semantics, steady-state characteristics, and heterogeneous modulation discrepancies into a shared high-dimensional manifold, enabling rapid adaptation with minimal labels and robust cross-domain generalization for high-quality multi-task EM signal modeling.

% challenge2：用打包的策略解决，详见xxx。（一个单独的标题能对应上）
% 针对电磁信号具有动态、长序列的特性，我们提出了一种支持可变长度输入的打包方法，如图Fig.~\ref{fig:model} (b). 所示。具体而言，我们首先将多条长度不一的 IQ 样本拼接打包为一条超长序列，从而实现高效的多长度联合训练。同时，该机制也支持在推理阶段对任意长度输入的灵活处理。方法的详细实现见第 x 节。
To handle the dynamic and long-sequence nature of EM signals, we propose a packing strategy that supports variable-length inputs, as illustrated in Fig.~\ref{fig:model}(b). Specifically, we first concatenate multiple IQ samples of varying lengths into a single ultra-long sequence, enabling efficient joint training across different lengths. This mechanism also allows flexible processing of inputs of any length during inference. Detailed implementation is provided in Section~\ref{packing}.

\subsection{IQ tokenizer}
%
%IQ tokenizer：IQ 离散token-patch-embedding
% 面向电磁信号种类繁多、场景各异、任务多样化，加之采样率、调制方式与信道条件高度异构的特性，我们提出一套面向 IQ 信号的“离散化-分词-嵌入”流水线：首先将连续而稀疏的 IQ 流均匀切分为局部 patch（如图~\ref{fig:pack}），在保留相位-幅度耦合的同时显著压缩序列长度；随后设计轻量级 IQ Tokenizer，以采样率自适应嵌入技术将每个 patch 映射为高维 IQ token；最终由专为 IQ 结构定制的一维时序 Transformer 聚合局部-全局上下文特征。
Faced with the highly heterogeneous nature of EM signals due to their varied types, diverse scenarios, multiple tasks, differing sampling rates, modulation schemes and channel conditions, we present a pipeline of discretization, tokenization and embedding. The continuous yet sparse IQ stream is divided evenly into local patches that preserve the phase amplitude coupling while greatly shortening the sequence length. An IQ tokenizer maps each patch into a high dimensional sample token through a embedding by concatenating the embedded sampling rate. Finally a one dimensional temporal Transformer, specifically designed for IQ structures, aggregates local and global contextual features.

% 采样率sensitive
% 采样率是 IQ 信号自带的元信息，跨数据集差异大、动态范围 kHz 到 MHz。我们为每个信号样本动态插入一个“采样率 token”，与常规 cls token 并排置于序列开头：二者角色独立，仅靠位置编码区分，cls token 负责全局聚合，采样率 token 始终可见，不加入掩码。于是模型在重建过程中自然看到“这条波形以何采样率采集”，对脉宽、PRI、带宽等采样率敏感的任务无需额外手段即可对齐。
Sampling rate is an intrinsic, stable meta-attribute that spans kHz–MHz across datasets. Each signal is prefixed with both a dedicated sampling-rate token and the standard classification token. The two are functionally isolated and differentiated solely by positional encodings: the classification token yields a global representation, whereas the sampling-rate token is left unmasked and always visible. During reconstruction, the model therefore continuously perceives the acquisition rate of each signal, naturally aligning itself with sampling-rate-sensitive tasks—such as pulse-width, PRI, or bandwidth regression—without further scaffolding.
\subsection{Low-Redundancy Length Adaptive Multi-Signal Packing}\label{packing}
\begin{figure*}[htbp]
  \centering
  \includegraphics[width=0.7\linewidth]{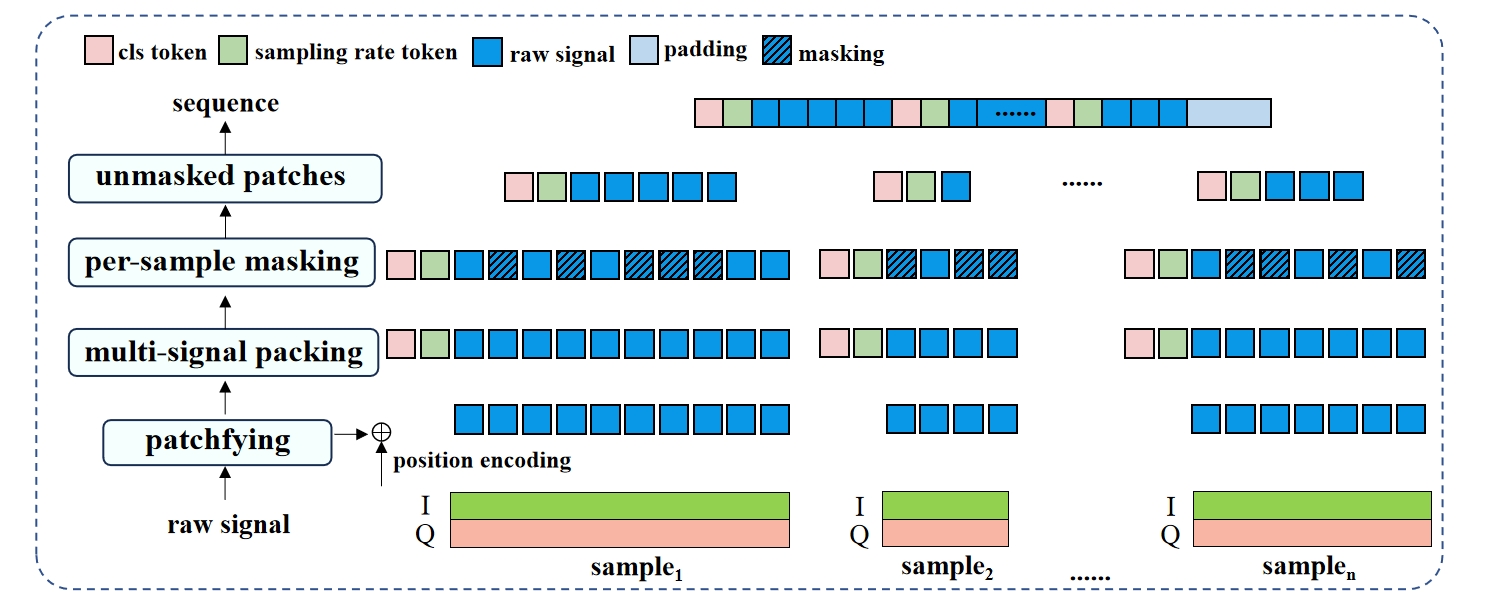}
  \caption{Low-redundancy length adaptive multi-signal packing and per-sample masking. Multiple signal samples are dynamically packed into fixed-length sequences, followed by per-sample masking strategy.}
  \label{fig:pack}
\end{figure*}
%
% 为实现对长短序列的自适应处理，兼容长序列且低冗余地统一训练，我们创新性地提出了可变长度信号打包策略，以替代传统的批量填充（padding）方法。我们的训练数据来自多个异构数据库，序列长度分布显著异质，从128到4,096不等，跨度达32倍。在这种高度非均匀的长度分布下，传统padding方法会引入大量无效零填充，显著增加显存占用和计算资源开销，严重制约训练效率和硬件利用率。为此，我们设计了一种动态打包机制：预设序列长度上限作为打包容量，按样本到达顺序依次填入，直至剩余容量不足时开启新序列。为确保模型能够准确区分并专注于各独立信号，防止信号间交叉干扰和信息混淆，系统还记录了每个样本在打包序列中的边界索引，保证信息完整且高效利用。相比传统的padding方法，信号打包有效克服了零填充导致的计算冗余和内存瓶颈问题，为大规模预训练提供了一个切实可行且高效的解决方案。
To enable adaptive processing of signals with varying lengths, and to support unified training with long samples in a memory-efficient manner, we propose a novel variable-length signal packing strategy as an alternative to conventional batch padding. Our training data is sourced from multiple heterogeneous datasets, exhibiting significant variation in signal lengths—from 128 to 4,096—spanning a 32× range. Under such highly non-uniform length distributions, traditional padding leads to excessive zero-padding, resulting in substantial memory overhead and computational inefficiency, which severely limits training scalability and hardware utilization.
To address this, we design a dynamic packing mechanism that predefines a maximum sequence length as the capacity for each packed sequence. Incoming IQ samples are sequentially inserted until the remaining capacity is insufficient, at which point a new sequence is initiated. To ensure that the model can accurately distinguish and focus on individual signals—preventing interference and information leakage across samples—we record the boundary indices of each sample within the packed sequence. This preserves information integrity and enables efficient signal utilization.
Compared with traditional padding-based approaches, our packing strategy effectively mitigates the computation redundancy and memory bottlenecks caused by zero-padding, providing a practical and scalable solution for large-scale pretraining.

% 基于预训练语料规模庞大的特点，即便采用信号打包技术，单个 epoch 仍需遍历海量样本。为此，我们设计了一套异步数据加载-训练协同系统：在 CPU 端开辟统一缓冲区，以生产者-消费者模型异步组织数据；CPU 负责并行加载、预处理与缓存管理，CPU与GPU通过流式调度实现紧密协作，形成并发异步流水线。这种设计使数据采集与模型训练交替进行，最大化内存利用率，凭借该高效架构，一个epoch内能够处理多达8100万条数据，有训练效率与资源利用率均获显著增益。
Given the large scale of the pre-training corpus, even with the multi-signal packing, each epoch must still process a massive number of samples. To address this, we design an asynchronous data loading and training coordination system. A unified buffer is allocated on the CPU side, where data is organized asynchronously based on a producer-consumer model. The CPU is responsible for parallel data loading, preprocessing, and buffer management, while a streaming scheduling mechanism enables tight coordination between the CPU and GPU. This forms a concurrent asynchronous pipeline that allows data acquisition and model training to proceed in parallel, thereby maximizing memory utilization. With this efficient architecture, we are able to process up to 81 million samples within a single epoch, resulting in significant improvements in training throughput and resource utilization.

% 此外，我们的网络结构设计具备良好的模块化与可扩展性。通过灵活替换下游任务头（task head），即可在共享的预训练基座模型上适配多种任务类型，包括分类、回归、检测、分离等不同任务形式。得益于我们的灵活打包方法，模型在推理阶段能够自然支持任意长度的信号输入，从而兼顾效率与输入多样性的需求。
Moreover, the proposed EMind is designed with strong modularity and scalability. By flexibly switching the downstream task head, the shared pre-trained foundation model can be adapted to various task types, including classification, regression, detection, and separation, as shwon in Fig.~\ref{fig:model} (c). Benefiting from our flexible packing method, the model naturally supports arbitrary-length signal inputs during inference, thereby balancing computational efficiency with input diversity.

\subsection{Per-Sample Masking in Multi-Signal Packing}

%针对电磁信号序列在打包后的掩码控制问题，我们面临一项重要挑战：现有掩码策略往往作用于整条打包后的长序列，由于序列中混合了多个长度不一的信号样本，无法精确区分样本边界，导致掩码率在整体序列上的表现呈现出较大浮动。例如，长度较短的信号可能会被完全掩码，而长度较长的信号则可能掩码不足，直接影响模型训练的稳定性和泛化能力，难以满足逐样本均衡掩码的需求。
We face a critical challenge in mask control for packed EM signal sequences: existing masking strategies typically operate on the entire concatenated long sequence. Since the sequence contains multiple signal samples of varying lengths mixed together, the precise boundaries of individual samples cannot be distinguished. As a result, the effective masking ratio fluctuates significantly across the overall sequence. For example, shorter signals may be fully masked while longer signals receive insufficient masking. This directly impacts the stability and generalization ability of model training, making it difficult to achieve balanced masking on a per-sample basis.

%为解决以上问题，我们设计并实现了一种高效的逐样本掩码算法。核心思路在于精确记录并利用每个信号样本的边界信息，对样本内部的数据进行独立随机化打乱和掩码采样。以往基于显式循环的实现方式虽然能够逐条样本施加掩码控制，但由于计算复杂度显著增加，破坏了流水线中序列打包策略的高效性。因此，我们创新性地提出了一种无需显式循环、基于矢量化操作的掩码控制机制。
To address the above issue, we designed and implemented an efficient per-sample masking algorithm. The core idea is to accurately record and utilize the boundary information of each signal sample, performing independent random shuffling and mask sampling within each sample. Although previous implementations based on explicit loops could apply mask control on a per-sample basis, they significantly increased computational complexity, undermining the efficiency of the pipeline’s sequence packing strategy. Therefore, we innovatively propose a masking control mechanism based on vectorized operations that requires no explicit looping.

%具体而言，对于每条长度为 l 的信号样本，我们首先在其序列开头插入一个代表采样率的编码 token，跟classification token拼到一起。随后，将该信号中每个采样点的位置归一化到区间 $[0,形成等距位置坐标集合{ \frac{1}{l-1}, \frac{2}{l-1}, \dots, 1 }$。
Specifically, for each signal sample of length $l$, we first insert a sampling rate encoding token at the beginning of the sample, concatenated together with the classification token. Next, we normalize the positions of each sampled point within the signal to the interval \([0, 1]\), forming an evenly spaced position coordinate set:
\begin{equation}
\left\{\frac{1}{l-1}, \frac{2}{l-1}, \ldots, 1\right\}.
\end{equation}
%
%借助这一标准化位置标识，我们对样本内部信号点进行随机顺序打乱。接下来，按照预定义的掩码率，从打乱后的样本数据中有选择性地选取有效的 Patch 供模型训练使用。此种做法有效保证了每条信号样本内部均匀且固定比例的内容被掩码，无论信号长度如何变化，掩码施加均能保持一致性和公平性，避免了掩码率浮动问题。
Using this normalized positional information, the signal points inside each sample are randomly shuffled. Then, according to the predefined masking ratio, a selective subset of valid patches is chosen from the shuffled sample data for model training. This approach effectively guarantees that a fixed proportion of content within every individual signal is masked evenly, ensuring consistent and fair mask application regardless of signal length variability, thereby avoiding mask ratio fluctuations.

% 值得一提的是，我们的掩码机制特别考虑了采样率 token 的特殊地位，保持其位置固定且始终参与训练。采样率 token 作为每条样本的固有元信息，不被掩码且不参与随机打乱，确保模型在训练过程中能够稳定获取信号采样率的关键信息，提升对信号时频特性的敏感度和识别能力。综上，基于上述机制，模型在训练时仅需对未被掩码的有效 patch 进行前向传播计算，极大地减小了冗余计算开销，显著提升整体计算效率和训练速度，如图~\ref{fig:pack} 所示。
It is worth noting that our masking scheme specially considers the unique status of the sampling rate token by keeping its position fixed and always included during training. As an intrinsic meta-information token for each sample, the sampling rate token is neither masked nor involved in the random shuffling, which ensures the model consistently captures key information about the signal’s sampling rate, enhancing sensitivity and recognition of its time-frequency characteristics during training. Based on the above mechanism, the model only needs to perform forward propagation on the unmasked valid patches during training, drastically reducing redundant computation and significantly improving overall computational efficiency and training speed, as illustrated in Figure~\ref{fig:pack}.

% 架构里面的loss
\subsection{Hardware-aware Dataset-adaptive Weighting}

% 我们提出的硬件可感知训练框架不仅为海量异构电磁信号的预训练提供了高效的算力支撑，也针对多源数据之间存在的难度差异大、收敛速度不一致等核心挑战，进一步设计了损失可监测的动态采样机制。具体而言，系统在初始化阶段以等权策略（即1:1:1:…）为各数据子集设置统一的采样起点，并在训练过程中持续监控每个子集的训练损失与验证损失变化。通过对各数据子集的损失曲线进行实时分析，系统可识别出训练进度缓慢或存在过拟合、遗忘等风险的子集。针对这类子集，系统将动态提升其采样权重，以增强模型对困难子集的学习能力；而对于易学习或出现性能退化的子集，则适当下调其采样频率，以降低过拟合风险，确保训练过程的稳定性与泛化性。
The proposed hardware-aware training framework not only provides efficient computational support for large-scale pre-training on heterogeneous EM signals, but also addresses key challenges such as significant difficulty variance across data sources and imbalanced convergence speeds. To this end, we further design a dataset-adaptive weighting mechanism. Specifically, the system initializes with an equal-weight sampling strategy (i.e., 1:1:1:…) across all datasets as the training baseline. During training, it continuously monitors the evolution of training and validation losses for each dataset. By analyzing the loss trajectories, the system identifies datasets that exhibit slow convergence or signs of performance degradation (e.g., overfitting or catastrophic forgetting). The sampling weights of such datasets are then increased dynamically to enhance the model’s learning on more challenging samples, while the sampling frequency of easily-learned or overfitting-prone datasets is reduced accordingly to mitigate overfitting and maintain training stability and generalization.

% 为实现上述策略，我们在数据缓冲区内部引入精确的配比控制机：借助内存映射的顺序访问策略，确保每个训练批次中，不同难度等级或类别的数据样本能够按照预设比例进行采样。具体实现上，采用内存映射的顺序访问策略，为每个数据子集分配独立的读取指针。该指针的移动速度根据目标采样比调整，从而保持每次训练中数据库采样比例的准确性。指针移动速度随最新权重实时调节，确保每个训练批次都能按照动态更新的比例精准混合样本。该机制既强化了对难学样本的关注，又抑制了对易过拟合数据的过度依赖。具体的各数据库预训练采样权重详见Section X。
To implement this strategy, we introduce a precise sampling weight control mechanism within the buffer. Leveraging a memory-mapped sequential access strategy, the system ensures that samples of varying difficulty levels or classes are drawn in each training batch according to the target sampling weights. Concretely, each dataset is assigned an independent pointer whose movement speed is adjusted based on the current target sampling weight. These speeds are updated in real time according to the latest weight assignments, ensuring that each batch contains a mixture of samples that reflects the dynamically adjusted proportions. This mechanism not only strengthens the model’s focus on hard-to-learn samples but also suppresses excessive reliance on subsets prone to overfitting. The used sampling weights for each dataset in this pre-training are detailed in Section~\ref{data_weight}.

\section{Experiments}
\begin{table*}[t]
  \caption{Multi-task Dataset}
  \label{tab:ft-data}
  \centering
  \resizebox{0.75\textwidth}{!}{
  \begin{threeparttable}
  \begin{tabular}{lcccccc}
    \toprule
     & Task & Number & Signal & Sampling  & SNR &  Data\\
     Dataset & Type* & of Types &  Length &  Rate (Hz) & Range (dB) & Source\\
    \midrule
    RML2016.10A~\cite{o2016radio} & AMC & 11 & 128 & 1M & [-20:18] &   - \\
    RML2016.10B~\cite{o2016radio} & AMC & 10 & 128 & 1M & [-20:18] &   - \\
    RML2016.04C~\cite{o2016convolutional} & AMC & 11 & 128 & 1M & [-20:18] &   - \\
    RML2018.01A~\cite{o2018over}  & AMC & 24 & 1024 & 1M &  [-20:30] &  - \\
    RadChar~\cite{huang2023multi} & RWC/RPE & 5 & 512 & 3.2M & [-20:20] & - \\
    % ORACLE~\cite{sankhe2019oracle} & RFFI & 16& 128 & 5M & - & - \\
    ADS-B~\cite{ya2022large} & RFFI & 198& 3000& 50M&  - & - \\
    % Hovering UAVs & RFFI & 7 & 200 & 10M &  - &  - \\
    \midrule
    \bf EM-AIS & RFFI & 112 & 3,840 &  156.25M & - &  Real-world \\
    \bf EM-Infer-Comm & WII & 9  &  1,024 & 20M  & \textit{[-20:20]} &  Simulation \\
    \bf EM-Infer-Radar & WII & 12 &  1,024 & 20M & \textit{[-20:20]} &  Simulation \\
    \bf EM-Radar-Mix & BSS & - & 1,024 & 5M & 12 &  Simulation \\
    \bf EM-Denoise-Signal & SD & - & 1,024 & 20M & [-3:20] & Simulation \\
    \bottomrule
  \end{tabular}
  \begin{tablenotes}
    \footnotesize
    \item[*] AMC represents Automatic Modulation Classification; RWC refers to Radar Waveform Classification; RPE stands for Radar Parameter Estimation; WII denotes Wireless Interference Identification,  RFFI indicates Radio Frequency Fingerprint Identification; SD refers to Signal Denoising; and BSS denotes Blind Source Separation.
    % \item[$\dagger$] ISR applies to interference datasets; SNR applies to all others.
  \end{tablenotes}
  \end{threeparttable}
  }
\end{table*}
\begin{table*}[t]
  \caption{Classification results compared with State-of-the-art. BOLD indicates the best performance.}
  \label{tab:cls}
  \centering
  \resizebox{1.\textwidth}{!}{
  \begin{threeparttable}
  \begin{tabular}{c|cccc|cc|cc}
    \toprule
    TASK & \multicolumn{4}{c|}{AMC} & \multicolumn{2}{c|}{RFFI}  & \multicolumn{2}{c}{WII} \\
    \midrule
    METHOD & RML2016.10A* & RML2016.10B* & RML2016.04C* & RML2018.01A & ADS-B & EM-AIS & EM-Infer-Comm & EM-Infer-Radar \\
    \midrule
    ResNet \cite{he2016deep} & 50.04 & 54.88 & 55.97 & 43.06 & 84.51 & 42.02 & 76.79 & 71.64 \\
    MCNet \cite{huynh2020mcnet} & 53.52 & 59.22 & 59.57 & - & - & - & - & - \\
    CNN2 \cite{o2018over} & 53.25 & 57.14 & 59.45 & - & - & - & - & - \\
    GRU2 \cite{hong2017automatic} & 58.80 & 64.11 & 63.13 & - & - & - & - & -  \\
    DAE \cite{ke2021real} & 58.97 & 61.46 & 55.91 & - & - & - & - & -   \\
    CGDNN \cite{njoku2021cgdnet} & 56.57 & 58.26 & 60.34 & - & - & -  & - & -  \\
    Transformer \cite{vaswani2017attention} & 59.27 & 63.10 & 65.41 & 59.45 & 78.77 & 39.12 & 80.44 & 77.05 \\
    MSNet \cite{zhang2021novel} & 58.33 & 63.49 & 63.66 & - & - & -  & - & - \\
    AMC\_Net \cite{zhang2023amc} & 59.10 & 63.38 & 63.01 & - & - & -  & - & -  \\
    SpectrumFM \cite{zhou2025spectrumfm} & 63.72 & 65.35 & 73.37 & - & - & -  & - & - \\
    \midrule
    EMind & 62.51 & \bf 65.45 & \bf 74.34 & \bf 63.83 & \bf 99.87 & \bf 57.07 & \bf 81.70 & \bf 79.19  \\
    \bottomrule
  \end{tabular}
  \begin{tablenotes}
  \footnotesize
  \item[*] As stated in the scikit-learn documentation ({\url{https://scikit-learn.org/stable/modules/generated/sklearn.metrics.recall_score.html}}), weighted recall is equivalent to accuracy in single-label multiclass settings. Therefore, we treat the recall reported in \cite{zhou2025spectrumfm} as overall accuracy (OA) for RML2016.10A, RML2016.10B, and RML2016.04C.
  \end{tablenotes}
  \end{threeparttable}
  }
\end{table*}
%
% 本章旨在全面验证我们提出的电磁基座模型在多样化电磁信号任务中的有效性与通用性。我们首先介绍预训练阶段的详细设置，包括数据设置、模型配置及优化参数。然后，我们围绕多个典型电磁信号任务开展下游评估实验，以系统检验模型在电磁特征建模、跨任务迁移以及小样本泛化等方面的表现。Automatic Modulation Classification (AMC), Wireless Technology Classification (WTC), Radar Waveform Classification (RWC), Radar Parameter Estimation (RPE), Wireless Interference Identification (WII), and Radio Frequency Fingerprint Identification (RFFI)—which also covers specific application-oriented identification tasks such as Automatic Identification System (AIS)-based ship identification, Automatic Dependent Surveillance-Broadcast (ADS-B) aircraft recognition, and Unmanned Aerial Vehicle (UAV) recognition, as well as fundamental perception tasks like Signal Denoising (SD) and Blind Source Separation (BSS). 
In this section, we aim to comprehensively validate the effectiveness and generalizability of the proposed EMind across a variety of EM signal tasks. We begin by detailing the pre-training setting, including training strategy and model specifications. Subsequently, we conduct downstream evaluation experiments on several representative tasks to systematically assess the model’s capability in EM feature representation, cross-task transferability, and few-shot generalization.

\subsection{Pre-training setting}

\subsubsection{dataset weight of training strategy}\label{data_weight}

% 在预训练过程中，我们将所有预训练数据库按照9:1的比例划分为训练集和验证集，并从中选择信噪比（SNR）大于6的数据用于预训练。为提高模型在不同数据集上的适应能力，我们提出了一种分库级的实时自监督重建损失跟踪机制。该机制通过在训练过程中动态调整采样比率，对易于过拟合的数据库降低采样比率，对难以学习的数据库提高采样比率，从而增强模型对训练中困难子集特征的建模能力，并提升整体泛化性能。最终，我们的预训练数据库采样比率设置为，EM-Comm : HisarMod2019.1 : Panoradio HF : RadarCommDataset : EM-RadarParaIQSim : EM-Radar : WiSig : Northeastern RF : POWDER : Transmitter Classification : LoRa RF Datasets : Mono Receiver : DroneRFa : EM-Infer-Radar-v2 = 1 : 0.5 : 1 : 1 : 0.5 : 0.5 : 1 : 1 : 1 : 1 : 1 : 1 : 0.5 : 0.5。最后我们在24小时内结束了10个epoch的训练，并取了第 3.6 个epoch最后最终发布的预训练权重。
During the pre-training process, all pre-training datasets are split into training and validation sets in a 9:1 ratio, with data having an SNR greater than 6 selected for pre-training. We propose a library-level real-time self-supervised reconstruction loss tracking mechanism, which dynamically adjusts the sampling ratio based on training iterations. This mechanism reduces the sampling ratio for datasets prone to overfitting and increases the ratio for harder-to-learn datasets, thereby enhancing the model's ability to model features from difficult subsets and improving its overall generalization performance. Ultimately, the sampling ratios for our pre-training datasets are set as follows, EM-Comm : HisarMod2019.1 : Panoradio HF : RadarCommDataset : EM-RadarParaIQSim : EM-Radar : WiSig : Northeastern RF : POWDER : Transmitter Classification : LoRa RF Datasets : Mono Receiver : DroneRFa : EM-Infer-Radar-v2 = 1 : 0.5 : 1 : 1 : 0.5 : 0.5 : 1 : 1 : 1 : 1 : 1 : 1 : 0.5 : 0.5. The pre-training process is completed within 24 hours over 10 epochs. The final pre-trained weights used for downstream tasks are passively selected from the checkpoint at the end of the 3.6-th epoch, due to its empirically observed training stability and generalization performance.

\subsubsection{Model specifications}
% 预训练实验的超参数配置如下：遮罩比例设为75\%，编码器12层，解码器8层；优化器采用AdamW，基础学习率1e-4，预热比例10\%；训练10个epoch，batch size 40，最大序列长度6000；仅使用信噪比高于6的样本，注意力机制选用eager attention，patch size设为8；最终模型参数量达1.1亿。
The hyperparameter configuration for pretraining is as follows: the mask ratio is set to 75 \%, the encoder has 12 layers and the decoder has 8 layers; the optimizer is AdamW with a base learning rate of 1e-4 and a warmup ratio of 10 \%; training is carried out for 10 epochs with a batch size of 40 and a maximum sequence length of 6 000; only samples whose signal-to-noise ratio exceeds 6 are used, eager attention is employed, the patch size is fixed at 8, and the final model reaches 110 million parameters.

% 我们发布了两个模型，EMind-base和EMind-large。

% IQ 数据均采用绝对幅值归一化处理。考虑到电磁信号的物理特性，其幅度通常直接反映了信号的强度和能量分布，因此使用绝对幅值归一化有助于有效保留这一关键特征，从而增强模型对信号物理属性的感知能力。其具体的归一化公式为：xxxxx。此外，回归参数也进行了归一化处理，特别是在处理小时间值（如雷达参数）时，归一化有助于提高回归任务的训练性能和收敛性。对于雷达参数，采用了最大最小归一化方法。
The IQ data are normalized using absolute magnitude normalization. Considering the physical characteristics of EM signals, the magnitude often directly reflects the signal’s strength and energy distribution. Applying absolute magnitude normalization helps preserve this critical feature, thereby enhancing the model’s ability to perceive the physical properties of the signal, which is $\hat{\mathbf{IQ}} =  \mathbf{IQ} / (\max(|\mathbf{IQ}|) $. The regression parameters are also normalized, particularly when dealing with small time values (such as radar parameters in $\mu$s), to improve the training performance and convergence of regression tasks. For radar parameters, min-max normalization is applied.

\subsection{Downstream tasks}
% 基础模型的一个核心特征在于其具备跨越预训练数据分布、泛化至多种下游任务的能力。为严格验证这一能力，我们选取了一组与预训练数据集完全独立的微调数据集，并围绕电磁应用场景设计了四类具有挑战性的下游评估任务，以全面反映模型在实际任务中的表现。如表 \ref{tab:ft-data} 所示，评估的具体任务包括自动调制分类（AMC）、无线技术分类（WTC）、雷达波形分类（RWC）、雷达参数估计（RPE）、射频指纹识别（RFFI），以及信号去噪（SD）与盲源分离（BSS）等关键任务，这些数据库的设置细节见appendix。
A core feature of the foundational model is its ability to generalize across pretraining data distributions and adapt to a wide range of downstream tasks. To evaluate this capability, we select a set of fine-tuning datasets that are completely independent from the pretraining datasets, and design four challenging downstream tasks to comprehensively assess the model’s performance in electromagnetic application scenarios. 
As shown in Table \ref{tab:ft-data}, the tasks include Automatic Modulation Classification (AMC), Radar Waveform Classification (RWC), Radar Parameter Estimation (RPE), Wireless Interference Identification (WII), Radio Frequency Fingerprint Identification (RFFI), as well as Signal Denoising (SD) and Blind Source Separation (BSS), with the details of these datasets provided in the Appendix.
% 在我们的通用模型设计中，所有下游任务（包括分类、回归与重建）在同一结构下统一建模，无需修改模型主体架构。在微调阶段，直接输入 IQ 样本，可在低至 -20 dB 信噪比、 6,000 以下任意长度下进行推理，使模型在处理不同长度样本时具备一致性和可扩展性，以全面考察模型在电磁信号上的泛化能力、适应性与鲁棒性。
In our model design, all downstream tasks (including classification, regression, and reconstruction) are unified within the same structure, without the need to modify the model's main architecture. During fine-tuning, IQ samples are input directly, allowing for inference at signal-to-noise ratios as low as -20 dB and with sequence lengths up to 6,000. This ensures consistency and scalability when handling samples of varying lengths, enabling a comprehensive evaluation of the model’s generalization, adaptability, and robustness in EM signals.

% 我们设计了三个实验轨道（track），分别对应不同的能力维度。首先，在全量微调（Full Fine-tuning）设置中，对模型的全部参数进行训练，以验证其作为通用模型在充分训练条件下的适应性与性能上限。其次，线性探测（Linear Probe）实验通过在冻结基座模型参数的基础上，仅训练一个线性分类器，从而评估模型在预训练阶段所学习到的通用特征表征能力，反映其作为特征提取器的有效性。第三，小样本学习（Few-shot Learning）则模拟现实中标注数据稀缺的场景，在每类样本数极少的情况下进行微调，以验证模型在低资源条件下的迁移能力与数据效率。
We design three experimental tracks, each corresponding to a different capability dimension of the model. First, in the full fine-tuning setting, all parameters of the model are updated during training to evaluate its adaptability and performance upper bound under fully supervised conditions, thereby validating its effectiveness as a general-purpose model. Second, the linear probe experiment freezes the parameters of the foundation model and trains only a linear classifier on top, in order to assess the general representation capability learned during pre-training and to evaluate the model's effectiveness as a feature extractor. Third, the few-shot Learning setting simulates real-world scenarios with limited labeled data, where the model is fine-tuned with only a few samples per class. This track is used to evaluate the model’s transferability and data efficiency under low-resource conditions.

\begin{figure*}[htbp]  % 使用[H]来强制图像出现在当前位置
    \centering
    \includegraphics[width=0.75\linewidth]{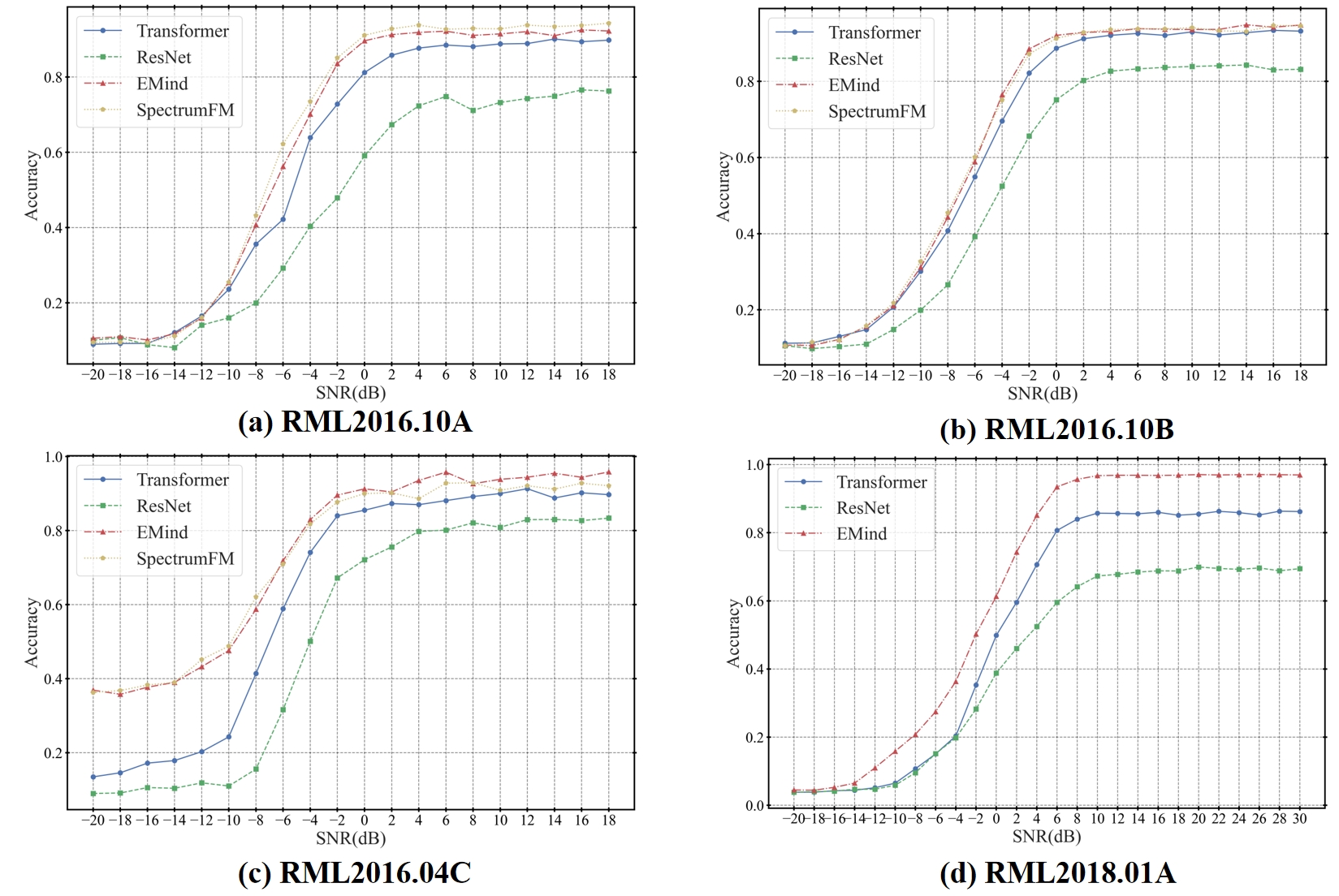}
    \caption{Performance comparison across different datasets in terms of accuracy at varying signal-to-noise ratios (SNR). (a)-(d) show the results for RML2016.10A, RML2016.10B, RML2016.04C, and RML2018.01A datasets, respectively.}
    \label{fig:snr}
\end{figure*}
%

% 为了更系统和细致地评估模型的表现，我们引入了多种评估指标，涵盖从分类、回归到盲源分离等不同任务维度的定量分析。在分类任务中，使用 Overall Accuracy（OA）来衡量整体分类性能，OA 表示模型预测正确的样本数占总样本数的比例，反映了模型在所有测试样本中的平均分类准确度，这是一种常用且直观的分类性能指标。回归任务中采用均方误差（MAE）作为基本指标，MAE 定义为预测值与真实值之间差异绝对值的平均值，能够有效衡量模型预测结果与真实值之间的偏差大小。对于盲源分离和信号去噪任务，我们使用均方误差（MSE）、信失真比（SDR）、信号干扰比（SIR）、伪影比（SAR）以及尺度不变信号失真比（SI-SDR）作为衡量标准，其中 MSE 反映分离信号与参考信号在时域上的整体重建精度，SDR 衡量目标信号功率与所有失真成分（包括噪声、干扰和伪影）能量之比，作为综合的分离质量指标，SIR 用于评估目标信号与其他源干扰的抑制能力，SAR 量化分离过程中引入的伪影程度，体现分离信号中伪像的比例，而 SI-SDR 则是 SDR 的尺度不变扩展，通过增益归一化消除幅度影响，提供更具鲁棒性的信号质量评价。在 few-shot 分类任务中，进一步引入 Kappa 系数来衡量分类结果的一致性并排除偶然因素，作为一种更严格的群体分类一致性评价指标，这些多样的评估指标共同构成了对模型在多任务、多维度场景下表现的全面量化依据。
To systematically and comprehensively evaluate the model's performance, we introduce multiple evaluation metrics covering quantitative analysis across different task dimensions, including classification, regression, and blind source separation. In classification tasks, we use Overall Accuracy (OA) to measure overall classification performance. OA represents the ratio of correctly predicted samples to the total number of samples, reflecting the average classification accuracy of the model across all test samples. It is a commonly used and intuitive metric for classification performance. In regression tasks, we adopt Mean Absolute Error (MAE) as the primary metric, which is defined as the average of the absolute differences between the predicted and actual values, effectively measuring the magnitude of the deviation between the model's predictions and the true values.
For blind source separation and signal denoising tasks, we use Signal Distortion Ratio (SDR), Signal Interference Ratio (SIR), Signal Artifacts Ratio (SAR), and Scale-Invariant Signal Distortion Ratio (SI-SDR) and Mean Squared Error (MSE) as the evaluation metrics. Among these, MSE reflects the overall reconstruction accuracy of the separated signal relative to the reference signal. SDR measures the ratio of the target signal power to the energy of all distortion components (including noise, interference, and artifacts), serving as a comprehensive indicator of separation quality. SIR evaluates the suppression of interference from other sources in the target signal, while SAR quantifies the degree of artifacts introduced during the separation process, representing the proportion of artifacts in the separated signal. SI-SDR, as a scale-invariant extension of SDR, normalizes by gain to eliminate amplitude effects, providing a more robust measure of signal quality.
In few-shot tasks, we further introduce the Kappa coefficient to assess the consistency of classification results and eliminate the influence of chance, serving as a more stringent metric for group classification consistency. These diverse evaluation metrics collectively provide a comprehensive quantitative foundation for evaluating the model's performance across multiple tasks and multi-dimensional scenarios.

\subsubsection{Classification}

% 在分类任务中，基座模型预训练提取出来的通用特征表示送入分类头，即线性分类器进行类别预测，并以交叉熵作为损失函数。我们评估了多个数据集，包括通信调制类型数据集 RML2016.10A~\cite{o2016radio}, RML2016.10B~\cite{o2016radio}, RML2016.04C~\cite{o2016convolutional}, RML2018.01A~\cite{o2018over},辐射源识别任务中的 ADS-B~\cite{ya2022large} 和 EM-AIS，以及干扰类型识别中的 EM-Infer-Comm 和 EM-Infer-Radar。其中，EM-AIS 是用于船舶辐射源识别任务的自采数据集，EM-Infer-Comm 和 EM-Infer-Radar是自建干扰仿真数据集。各数据集的具体信号长度如下：RML2016.10A、RML2016.10B、RML2016.04C 的信号长度为 128，RML2018.01A 为 1,024，ADS-B 为 3,000，EM-Infer-Comm 和 EM-Infer-Radar 也为 1,024。由于 EM-AIS 信号的原始序列较长11,520，为了减轻计算负担，我们根据其带宽特性并参考奈奎斯特采样准则，对其进行了三倍降采样，最终得到的信号长度为 3,840。关于数据集的划分，RML2016.10A、RML2016.10B、RML2016.04C 和 RML2018.01A 均采用 80% 的数据用于微调训练，ADS-B，EM-Infer-Comm 和 EM-Infer-Radar 采用 1:9 的训练与测试比例，而由于 EM-AIS 数据总量较小且类别较多，为确保训练集能够覆盖所有类别，我们采用了 5:5 的训练设置。
In the classification task, the general feature representations extracted from the pretrained base model are fed into a linear classifier for category prediction, with cross-entropy adopted as the loss function. We evaluate performance across multiple datasets, including the communication modulation recognition datasets RML2016.10A~\cite{o2016radio}, RML2016.10B~\cite{o2016radio}, RML2016.04C~\cite{o2016convolutional}, and RML2018.01A~\cite{o2018over}; the emitter identification datasets ADS-B~\cite{ya2022large} and EM-AIS; as well as the interference type recognition datasets EM-Infer-Comm and EM-Infer-Radar. Among them, EM-AIS is a proprietary dataset collected for shipborne emitter identification, while EM-Infer-Comm and EM-Infer-Radar are self-constructed synthetic datasets for interference classification. The signal lengths of these datasets are as follows: 128 for RML2016.10A, RML2016.10B, and RML2016.04C; 1,024 for RML2018.01A, EM-Infer-Comm, and EM-Infer-Radar; and 3,000 for ADS-B. Since the raw sequence length of EM-AIS is 11,520, to reduce computational overhead we perform threefold downsampling based on its bandwidth characteristics and the Nyquist sampling criterion, yielding a final signal length of 3,840. For dataset partitioning, we use 80\% of the data for fine-tuning on RML2016.10A, RML2016.10B, RML2016.04C, and RML2018.01A; a 1:9 training-to-testing split for ADS-B, EM-Infer-Comm, and EM-Infer-Radar; and a 5:5 split for EM-AIS, as its limited data volume and large number of categories necessitate ensuring coverage of all classes in the training set.

\paragraph{Automatic Modulation Classification and Radio Frequency Fingerprint Identification}
% \FloatBarrier
%
\begin{figure*}[htbp] 
    \centering
    \includegraphics[width=\linewidth]{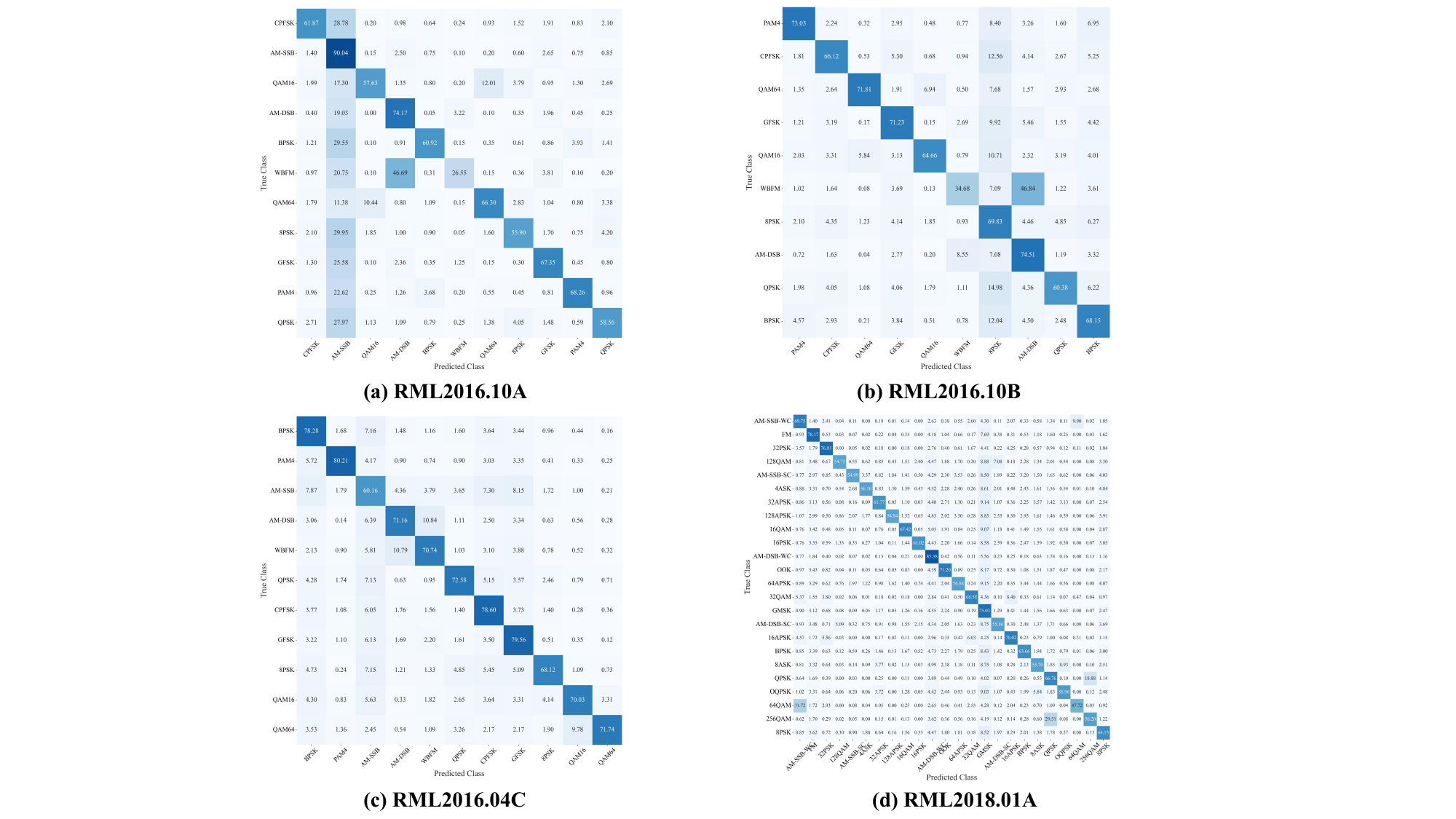}
    \caption{Confusion matrices for EMind on different datasets. (a)-(d) show the confusion matrices for RML2016.10A, RML2016.10B, RML2016.04C, and RML2018.01A, respectively, highlighting the model's performance across various modulation categories.}
    \label{fig:confusion}
\end{figure*}
%
% 我们的分类实验结果如表\ref{tab:cls}所示，其中RML2016.10A~\cite{o2016radio}、RML2016.10B~\cite{o2016radio}、RML2016.04C~\cite{o2016convolutional}采用了SpectrumFM \cite{zhou2025spectrumfm}的对比算法设置；而在RML2018.01A~\cite{o2018over}、ADS-B~\cite{ya2022large}、EM-Infer-Comm，EM-Infer-Radar 和 EM-AIS数据集上的对比算法则是我们基于ResNet \cite{he2016deep}和Transformer \cite{vaswani2017attention}模型的复现。
The classification experiment results are shown in Table~\ref{tab:cls}, where RML2016.10A~\cite{o2016radio}, RML2016.10B~\cite{o2016radio}, and RML2016.04C~\cite{o2016convolutional} use the contrastive algorithm setup from SpectrumFM~\cite{zhou2025spectrumfm}; while the comparison algorithms for the RML2018.01A~\cite{o2018over}, ADS-B~\cite{ya2022large} ,EM-Infer-Comm, EM-Infer-Radar and EM-AIS datasets are reproduced based on the ResNet~\cite{he2016deep} and Transformer \cite{vaswani2017attention} models. 
% 实验结果表明，我们的模型在调制分类、辐射源识别、干扰类型识别三个任务中均达到了SOTA性能。在调制分类任务中，EMind 在 RML2016.10B 和 RML2016.04C 上分别实现了 65.45% 和 74.34% 的准确率，均优于所有对比方法，其中在 RML2016.04C 上较 SpectrumFM 提升了 0.97 个百分点。在辐射源识别任务中，EMind 在 ADS-B 数据集上取得了 99.87% 的准确率，在更具挑战性的 EM-AIS 数据集上也达到 57.07%，显著领先于 Transformer 基准模型（39.12%），验证了方法在复杂环境下的鲁棒性。在干扰类型识别任务中，EMind 在 EM-Infer-Comm 和 EM-Infer-Radar 上分别获得 81.70% 和 79.19% 的准确率，均超越现有方法，进一步证明了所提出模型在高难度信号识别场景中的优越性能。
The experimental results demonstrate that our model achieves state-of-the-art performance across Automatic Modulation Classification (AMC), Radio Frequency Fingerprint Identification (RFFI), and Wireless Interference Identification (WII) tasks. In AMC, EMind attains accuracies of 65.45\% and 74.34\% on RML2016.10B and RML2016.04C, respectively, surpassing all competing approaches, with a 0.97 percentage point improvement over SpectrumFM~\cite{zhou2025spectrumfm} on RML2016.04C. In RFFI, EMind achieves 99.87\% accuracy on the ADS-B dataset, and further delivers 57.07\% on the more challenging EM-AIS dataset, substantially outperforming the Transformer baseline (39.12\%) and verifying the robustness of the proposed method under complex environments. For WII, EMind obtains 81.70\% and 79.19\% accuracy on EM-Infer-Comm and EM-Infer-Radar, respectively, outperforming all existing methods and further demonstrating the superior capability of our model in highly challenging signal recognition scenarios.

%图\ref{fig:snr}和图\ref{fig:confusion}展示了多个通信数据集在不同信噪比水平下的性能分析和混淆矩阵。图\ref{fig:snr}展示了在RML2016a、RML2016b、RML2016c和RML2018a数据集上，EMind与其他模型在不同信噪比条件下的总体准确度（OA）对比；图\ref{fig:confusion}展示了EMind在相应数据集上的混淆矩阵，详细呈现了其在各种调制类别上的分类性能，突出其优势以及潜在的误分类领域。
Figure~\ref{fig:snr} and Figure~\ref{fig:confusion} present performance analysis at different signal-to-noise (SNR) ratios and confusion matrices for multiple communication datasets. Figure~\ref{fig:snr} compares the overall accuracy (OA) of EMind with other models under various SNR conditions; Figure~\ref{fig:confusion} shows the confusion matrix for EMind, providing a detailed view of its classification performance across various modulation categories, highlighting both strengths and potential misclassification areas.

\paragraph{Feature-based Approach}

% 网络结构，数据库是什么为什么用，setting，信号长度，结果。

% 为验证预训练模型是否习得任务无关的通用表征，除全量微调外，我们还采用线性探针（linear probe）策略开展调制类型分类实验：冻结全部预训练权重，仅训练附加在输出特征上的线性分类器。该策略以最小化参数更新幅度，评估特征的线性可分性与跨任务泛化力，从而更纯粹地衡量预训练表征的质量。相比全量微调，线性探针计算开销更低，且能直接反映预训练特征本身的通用迁移能力。
To verify whether the pre-trained model has learned task-agnostic general representations, modulation classification experiments are conducted using a linear probing strategy in addition to full fine-tuning. In this setting, all pre-trained weights are frozen, and only a linear classifier appended to the output features is trained. This strategy minimizes the extent of parameter updates and is used to assess the linear separability and cross-task generalizability of the features, thereby providing a purer measure of the quality of the pre-trained representations. Compared to full fine-tuning, linear probing incurs lower computational overhead and directly reflects the inherent transferability of the pre-trained features.

\begin{figure}[htbp]
    \centering
    \includegraphics[width=\linewidth]{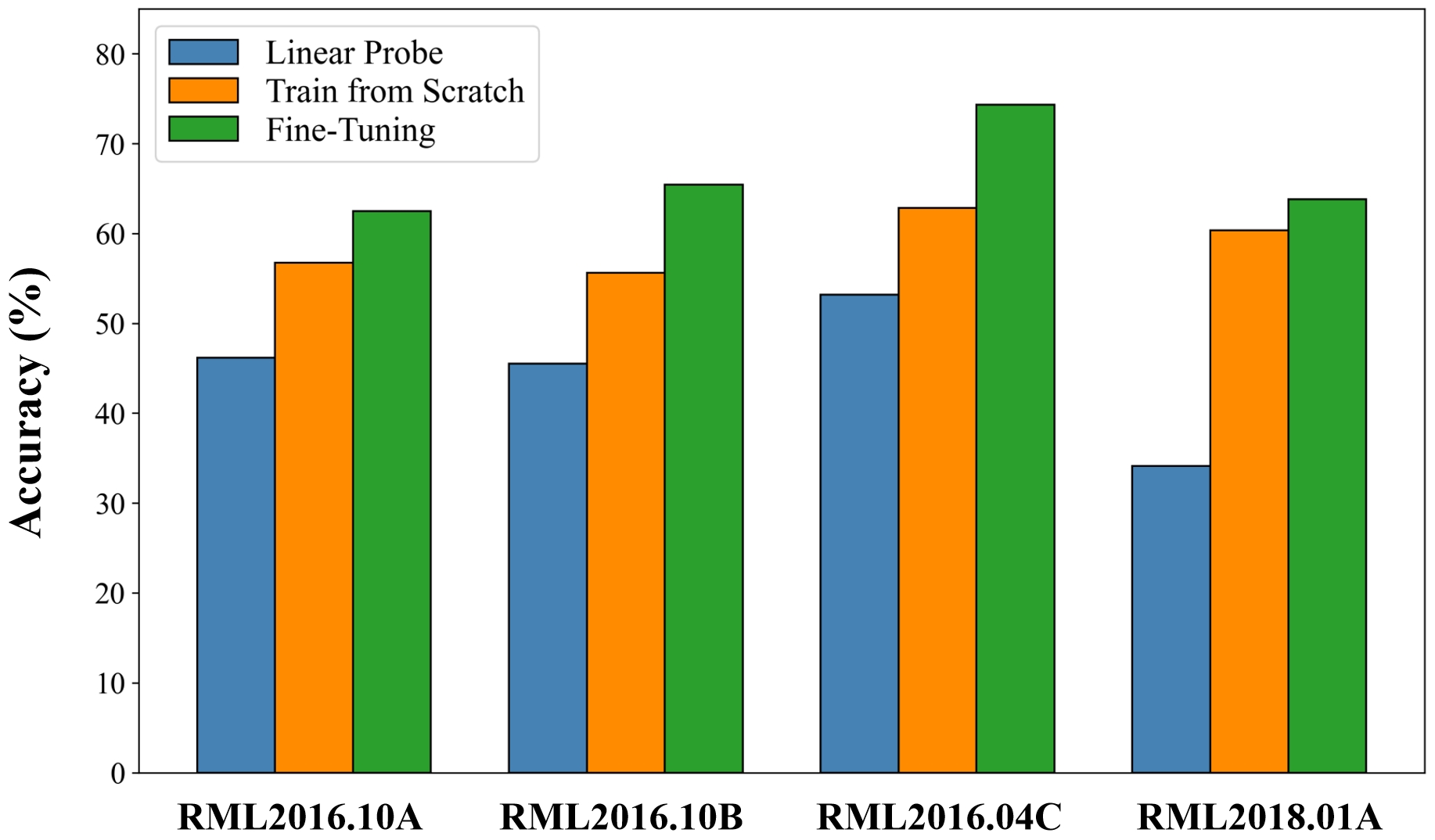}
    \caption{Classification accuracy on four public datasets RML2016.10A, RML2016.10B, RML2016.04C, and RML2018.01A under three training strategies: training from scratch, linear probing, and full fine-tuning.}
    \label{fig:linear}
\end{figure}

\begin{figure*}[htbp]
    \centering
    \includegraphics[width=0.8\linewidth]{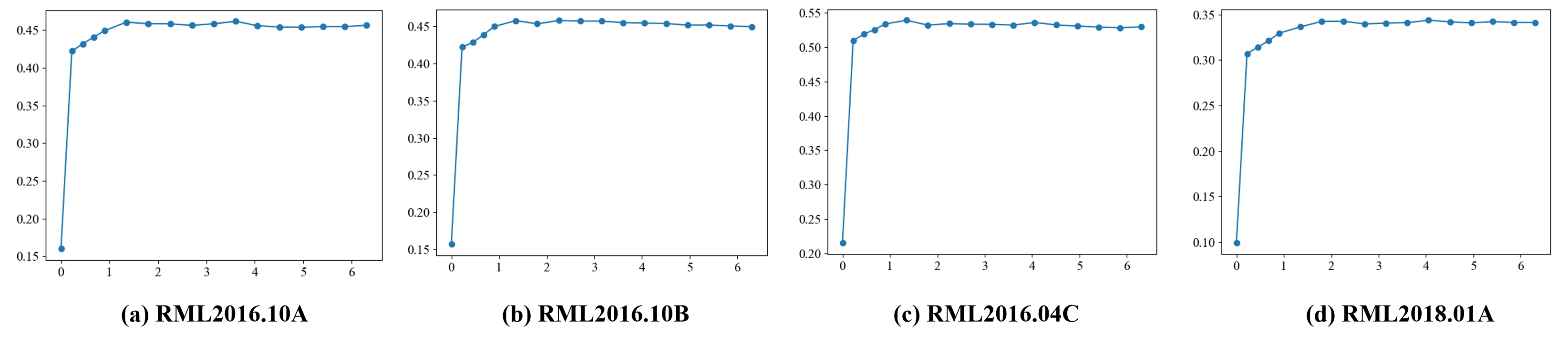}
    \caption{Evaluation of linear probing accuracy on checkpoints from different pre-training epochs. Rapid saturation of performance indicates that transferable and linearly separable features are learned early during pre-training.}
    \label{fig:linear_conv}
\end{figure*}

% 如图{fig:linear}所示，我们选取 RML2016.10A~\cite{o2016radio}、RML2016.10B~\cite{o2016radio}、RML2016.04C~\cite{o2016convolutional} 与 RML2018.01A~\cite{o2018over} 四个公开数据集，分别开展“从零训练”“全量微调”与“线性探针”三类实验。结果显示，仅冻结预训练权重、训练一层线性分类器，即可获得与全量微调相近的精度，充分证明模型表征的通用性与迁移能力。进一步，如图{fig:linear_conv}所示，我们对不同训练轮次的 checkpoint 进行线性探针评估。可见，随着预训练轮次增加，分类性能迅速上升，并在早期阶段即趋于饱和。这一现象表明，模型在训练伊始便已习得高质量、线性可分的通用特征；即便以早期权重作为固定特征提取器，也能显著加速下游任务收敛并提升最终性能。
As shown in Figure~\ref{fig:linear}, we evaluate our model on four public datasets including RML2016.10A~\cite{o2016radio}, RML2016.10B~\cite{o2016radio}, RML2016.04C~\cite{o2016convolutional}, and RML2018.01A~\cite{o2018over}, under three experimental settings: training from scratch, full fine-tuning, and linear probing. The results show that comparable accuracy to full fine-tuning is achieved by training only a single linear classifier on top of the frozen pre-trained features, demonstrating the generality and transferability of the learned representations. Furthermore, as illustrated in Figure~\ref{fig:linear_conv}, linear probing is performed on checkpoints from different pre-training epochs. It is observed that classification performance improves rapidly with increased pre-training duration and saturates at an early stage. This phenomenon indicates that the model acquires high-quality, linearly separable, and generalizable features early in training. Even when early-stage weights are used as fixed feature extractors, convergence of downstream tasks is significantly accelerated and final performance is enhanced.

\FloatBarrier

\subsubsection{Radar parameters estimation tasks}

\begin{table*}[htbp]
  \caption{Multi-task of RWC and RPE on RadChar~\cite{huang2023multi}. BOLD indicates the best performance. Lower MAE indicates better regression performance, while higher accuracy reflects better classification.}
  \label{tab:reg}
  \centering
  \resizebox{\textwidth}{!}{
  \begin{threeparttable}
  \begin{tabular}{cccccc}
    \toprule
    Methods & MAE $n_p$ (µs) & MAE $t_{pw}$ (µs) & MAE $t_{pri}$ (µs) & MAE $t_d$ (µs) & Class Accuracy (\%) \\
    \midrule
    & -10db 0db 10db all$^{\dagger}$ & -10db 0db 10db all & -10db 0db 10db all & -10db 0db 10db all & -10db 0db 10db all \\
    \midrule
    CNN1D & 0.729, 0.193, 0.085 - & 1.413, 0.560, 0.340 - & 0.999, 0.330, 0.209 - & 1.349, 0.385, 0.206 - & 75.7, 99.8, 100 - \\
    CNN2D & 0.793, 0.174, 0.090 - & 1.466, 0.801, 0.505 - & 1.054, 0.420, 0.299 - & 1.729, 0.638, 0.443 - & 67.3, 98.3, 99.8 - \\
    IQST-S & 0.733, 0.294, 0.251 - & 1.282, 0.628, 0.364 - & 0.816, 0.273, 0.192 - & 1.229, 0.415, 0.277 - & 79.2, 99.9, 100 - \\
    IQST-L & 0.752, 0.195, 0.124 - & 1.253, 0.579, 0.334 - & 0.799, 0.286, 0.225 - & 1.253, 0.379, 0.233 - & 79.1, 99.8, 100 - \\
    \midrule
    \bf EMind & \bf 0.330, 0.006, 0.005, 0.114 & \bf 0.797, 0.197, 0.080, 0.305 & \bf 0.463, 0.109, 0.085, 0.221 & \bf  0.708, 0.149, 0.092, 0.323 & \bf  86.65, 100, 100, 88.49 \\
    \bottomrule
    \end{tabular}
    \begin{tablenotes}
      \item[${\dagger}$] all indicates that the results are evaluated across the entire SNR range  from -20 dB to 20 dB.
      \end{tablenotes}
      \end{threeparttable}
  }
\end{table*}

% 考虑加个curve？

% 对于回归任务，回归头输出连续值，并采用均方误差等回归型损失函数进行优化。在该任务下，我们实现了多任务联合回归，将回归头扩展为多输出结构，以支持多个连续变量的联合建模，并与分类任务共同训练。我们的实验基于雷达数据集 RadChar~\cite{huang2023multi} 进行评估，同时对雷达波型进行分类，并对脉冲数（$n_p$）、脉宽（$t_{pw}$）、脉冲重复周期（$t_{pri}$）和脉冲时间延迟（$t_d$）等四个参数进行回归。具体而言，脉冲数的取值范围为 2 至 6，脉宽范围为 10 至 16 µs，脉冲重复周期（PRI）为 17 至 23 µs，脉冲时间延迟为 1 至 10 µs。为了公平对比，我们将数据集的划分设置为 70:15:15，以与 \cite{huang2023multi} 对齐，且信号长度为 512。结果如表 \ref{tab:reg} 所示，分类精度用oa衡量，回归精度用 MAE衡量，我们分别测试了测试集在 -10 db，0 db， 10 db信噪比下的性能，并测试了全部信噪比 -20db到20db下的性能。我们的联合分类回归性能大幅超过了sota水平。
For the regression task, the regression head outputs continuous values and is optimized using regression loss functions such as mean absolute error (MAE). In this task, multi-task joint regression is implemented by extending the regression head into a multi-output structure to support joint modeling of multiple continuous variables, which is trained jointly with the classification task. The experiments are evaluated on the radar dataset RadChar~\cite{huang2023multi}, where radar waveform classification is performed, and four parameters—pulse number ($n_p$), pulse width ($t_{pw}$), pulse repetition interval ($t_{pri}$), and pulse time delay ($t_d$)—are regressed. Specifically, the pulse number ranges from 2 to 6, pulse width ranges from 10 to 16 µs, pulse repetition interval (PRI) ranges from 17 to 23 µs, and pulse time delay ranges from 1 to 10 µs. To ensure a fair comparison, the dataset split is set as 70:15:15, following~\cite{huang2023multi}, with a signal length of 512. The results are shown in Table~\ref{tab:reg}, where classification accuracy is measured using overall accuracy (OA) and regression performance is measured using MAE. We test the performance on the test set at SNR of -10 dB, 0 dB, and 10 dB respectively, and also evaluate performance across all SNR range from -20 dB to 20 dB. Our joint classification-regression performance substantially outperforms the state-of-the-art (SOTA) methods.

\subsubsection{Blind source separation and Signal Denoise}

% 盲源分离（Blind Source Separation, BSS）是复杂电磁环境中极具挑战性的逆问题，旨在在不事先了解源信号特性或混合过程的情况下，从观测到的混合信号中恢复原始且相互独立的源信号。由于只能观测到由未知数量辐射源经未知信道混合后的带噪IQ波形，既无法预先获知源的真实个数，也无法在训练阶段获得任何干净的地面真值源信号，该问题属于病态问题，若无额外假设，可能不存在唯一解，并且系统对输入的微小扰动极为敏感，尤其在带噪IQ信号环境下。
Blind Source Separation (BSS) is a highly challenging inverse problem in complex electromagnetic environments. Its goal is to recover the original, mutually independent source signals from observed mixed signals without prior knowledge of the source signal characteristics or the mixing process. Since only noisy IQ signals can be observed, which are from the mixing of an unknown number of radiation sources through unknown channels, the true number of sources cannot be known in advance, and no clean ground truth source signals are available during training. This problem is ill-posed, without additional assumptions a unique solution may not exist, and the system is highly sensitive to small perturbations in the input, especially in noisy IQ signal environments.
% 
% 自动编码器（Autoencoder）因其能够从混合信号中学习紧凑且可区分的潜在表示而被广泛应用于盲源分离任务。通过将混合信号映射到低维潜在空间，自动编码器能够在无监督条件下分离潜在源信号。为适应源分离这一任务，我们实现了基于自编码器（AE）架构的微调框架，在预训练模型基础上引入多层线性压缩层，并通过限制解码器容量，强制模型从复杂的混合信号中提取简洁且具区分性的潜在特征。具体而言，如表{tab:radar_mix}所示，线性压缩层按照预设隐藏维度逐层作用，最终通过最后一层线性映射将特征压缩投影至固定的 $K$（在我们的设定中，$K=2$）个通道，每个通道以16维向量表示，以降低信号的表达复杂度。核心损失函数由两部分组成：其一是重建误差损失，支持对整体混合信号或分离后的各独立信号通道进行重建，后者通过排列不变训练策略（Permutation Invariant Training）解决信号顺序不确定性问题，确保训练目标与实际分离效果高度一致；其二是用于正则化潜在表示的 $\ell_2$ 范数约束项，鼓励编码特征保持稳定、稀疏并具有良好的判别性，从而提升表示的泛化能力和鲁棒性。这种联合损失设计不仅优化了信号的重建精度，还促进模型学习到稳定且具判别性的潜在表示。
Autoencoders are widely used in blind source separation tasks due to their ability to learn compact and discriminative latent representations from mixed signals. By mapping the mixed signals to a lower-dimensional latent space, autoencoders can separate the latent source signals in an unsupervised manner. To address the source separation task, a fine-tuning framework based on the Autoencoder (AE) architecture is implemented. A multi-layer linear compression layer is introduced on top of the pretrained model, and by constraining the decoder capacity, the model is forced to extract compact and discriminative latent features from the complex mixed signals. Specifically, as shown in Table~\ref{tab:radar_mix}, the linear compression layers are applied progressively according to predefined hidden dimensions, and the features are ultimately compressed and projected to a fixed number of $K$ channels (with $K=2$ in our setting) through the final linear mapping, where each channel is represented by a 16-dimensional vector, reducing the complexity of signal representation. The core loss function consists of two components: the first is the reconstruction error loss, which supports the reconstruction of either the entire mixed signal or the separated independent signal channels. The latter addresses the signal order uncertainty using a permutation invariant training strategy, ensuring a high alignment between the training objective and the actual separation performance. The second component is the $\ell_2$-norm regularization term applied to the latent representations, which encourages the encoded features to remain stable, sparse, and discriminative, thus improving the generalization and robustness of the representation. This joint loss design not only optimizes the reconstruction accuracy of the signal but also promotes the model to learn stable and discriminative latent representations.
\begin{table}[h!]
  \caption{Model layers architecture for the downstream task of blind source separation. Note: K = 2.}
  \label{tab:radar_mix}
  \centering
  \resizebox{0.5\textwidth}{!}{
  \begin{tabular}{lcccccc}
    \toprule
    INPUT & Layer 1 & Layer 2 & Layer 3  & Layer 4 & OUTPUT \\
    \midrule
    $D_{Enc} \times num_{patch}$ & 4096 & 2048 & 1536 & 1024 & $16K$ \\
    \bottomrule
  \end{tabular}
  }
\end{table}

\begin{table}[htbp]
  \caption{Performance Comparison of Blind Source Separation (BSS) on EM-Radar-Mix.}
  \label{tab:bss}
  \centering
  \resizebox{0.47\textwidth}{!}{
  \begin{tabular}{cccccc}
    \toprule
    Setting   & SDR & SIR & SAR & SI-SDR & MSE \\
    \midrule
     Linear prob  & 4.85 & 10.73 & 6.92 & -5.88 & -13.85 \\
     Fine-tune  & 5.74 & 11.60 & 7.83 & -3.85 & -15.32 \\
    \bottomrule
  \end{tabular}
  }
\end{table}

\begin{figure*}[htbp]
    \centering
    \includegraphics[width=1.\linewidth]{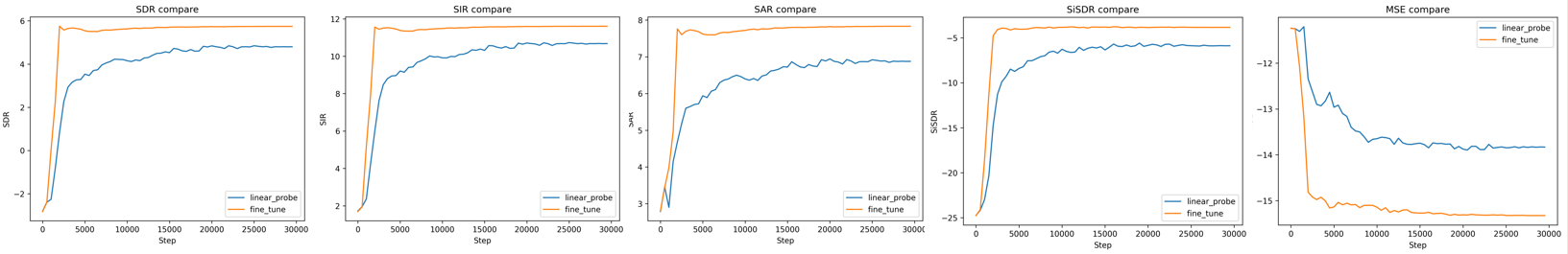}
    \caption{The BSS Eval comprises Signal-to-Distortion Ratio (SDR), Signal-to-Interference Ratio (SIR), Signal-to-Artifacts Ratio (SAR), Scale-Invariant Signal-to-Distortion Ratio (SI-SDR), and Mean Squared Error (MSE), where higher values of the first four indicate superior separation quality, whereas lower MSE signifies higher fidelity.}
    \label{fig:bss_metrics}
\end{figure*}

\begin{figure*}[htbp]
  \centering
  \includegraphics[width=1.0\linewidth]{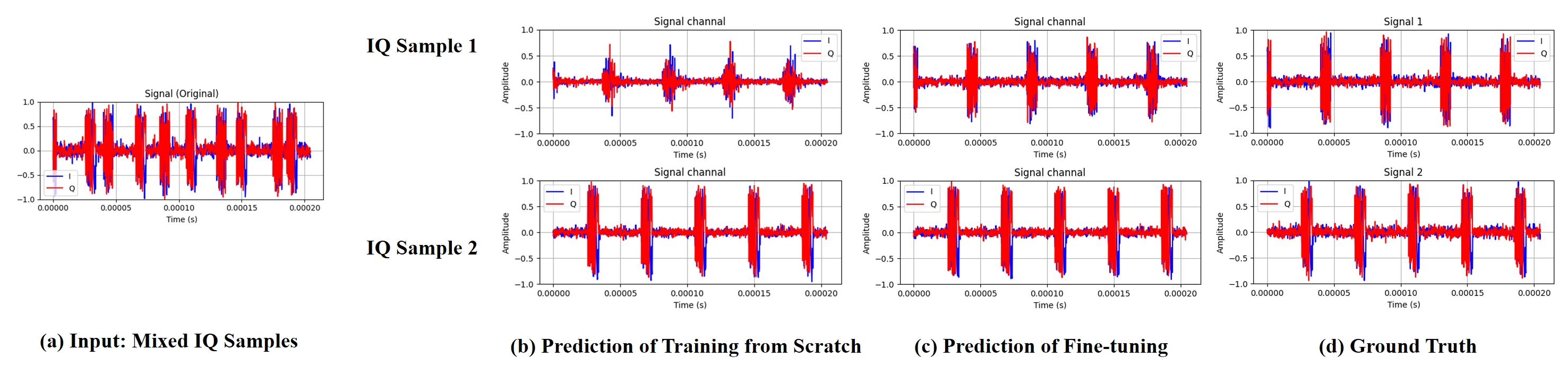}
  \caption{Visualization of IQ signal blind source separation (BSS) results. (a) mixed IQ signal input, (b) separation results of training from scratch, (c) separation results with fine-tuning, (d) ground truth.}
  \label{fig:bss_vis}
\end{figure*}

\begin{figure*}[htbp]
  \centering
  \includegraphics[width=0.65\linewidth]{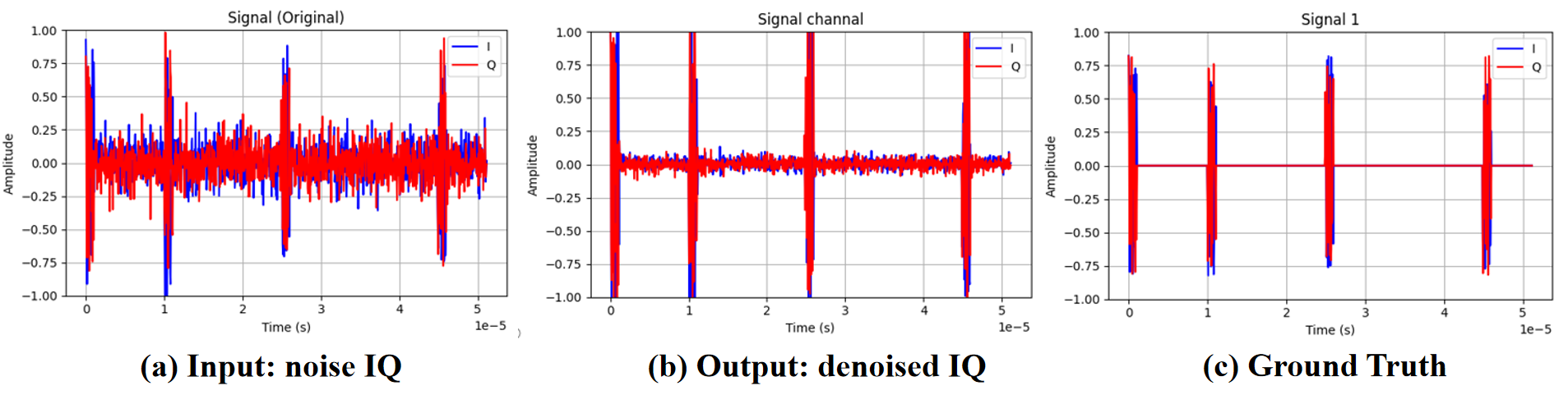}
  \caption{Visualization of IQ signal denoising. (a) noisy IQ input. (b) denoised prediction. (c) ground truth. The comparison illustrates the model's effectiveness in removing noise, achieving excellent consistency in amplitude and phase alignment.}
  \label{fig:denoise_vis}
\end{figure*}

\begin{table*}[t]
  \caption{Comparison of Methods for few shot on classification tasks. BOLD indicates the best performance.}
  \label{tab:few}
  \centering
  \resizebox{0.85\textwidth}{!}{
  \begin{tabular}{c|cccc|cccccc}
    \toprule
    TASK & \multicolumn{4}{c|}{AMC} & \multicolumn{6}{c}{RWC} \\
    \midrule
    \multirow{3}{*}{METHOD} & \multicolumn{4}{c|}{RML2016.10A} & \multicolumn{6}{c}{RadChar} \\
    \cline{2-11}
    & \multicolumn{2}{c|}{50 shot}  & \multicolumn{2}{c|}{100 shot} & \multicolumn{2}{c|}{10 shot}  & \multicolumn{2}{c|}{50 shot} & \multicolumn{2}{c}{100 shot}  \\
    \midrule
    &  OA(\%) &  Kappa  &  OA(\%) &  Kappa  &  OA(\%) &  Kappa  &  OA(\%) &  Kappa  &  OA(\%) &  Kappa \\
    \midrule
    ResNet & 38.97 & 32.87 & 42.01 & 36.21 & 71.94 & 64.93 & 79.89 & 74.86 & 80.97 & 76.21 \\
    Transformer & 38.40 & 32.25&	39.81&	33.79&	64.27&	55.37&	76.42&	70.52&	79.21&	74.01\\
    \midrule
    EMind & \bf 48.38 & \bf 42.85 & \bf 50.10 & \bf 45.12 & \bf 78.37 & \bf 72.97 & \bf 81.94 & \bf 77.43 & \bf 83.14 & \bf 78.93 \\
    \bottomrule
  \end{tabular}
  }
\end{table*}

% 我们使用自建数据集EM-Radar-Mix对盲源分离任务进行评估。由于该任务的复杂性与创新性，目前市面上缺乏适用的公开数据集，因此我们构建了这一具有挑战性的数据集。EM-Radar-Mix数据集包含123,200条样本，其中训练集、验证集和测试集的样本数量分别为100,800、11,200和11,200，信号长度为1,024。该数据集包含8种原始雷达信号，这些信号在采样率为5 MHz、信噪比（SNR）设定为12 dB的条件下进行混合。在样本构建过程中，这些信号通过两两组合或单独出现的方式进行混合，每条样本中所混合的具体信号类别对模型是未知的，从而形成典型的盲源分离任务场景。
The blind source separation task is evaluated on the self-constructed EM-Radar-Mix dataset. Due to the complexity and innovation of this task, there is currently a lack of suitable publicly available datasets, prompting the construction of this challenging dataset. The EM-Radar-Mix dataset consists of 123,200 samples, with 100,800 samples for the training set, 11,200 samples for the validation set, and 11,200 samples for the test set. The signal length is 1,024. The dataset contains 8 types of original radar signals, which are mixed under conditions of a 5 MHz sampling rate and a signal-to-noise ratio (SNR) set to 12 dB. During sample construction, these signals are mixed either in pairs or individually, and the specific signal types mixed in each sample are unknown to the model, thereby creating a typical blind source separation task scenario.

% 本研究采用BSS Eval工具包中定义的一系列标准化评估指标，对测试集中随机抽取的100个信号样本进行了性能评估。所选评估指标包括信号失真比（SDR）、信号干扰比（SIR）、信号伪影比（SAR）、尺度不变信号失真比（SI-SDR）以及均方误差（MSE）。这些指标是盲源分离（BSS）领域广泛认可的性能衡量标准，能够全面反映模型在信号重建过程中的失真水平、干扰抑制效能及分离精度。
A series of standardized evaluation metrics defined in the BSS Eval toolkit \cite{vincent2006performance} are used to assess the performance of 100 randomly sampled signal samples from the test set. The selected evaluation metrics include Signal-to-Distortion Ratio (SDR), Signal-to-Interference Ratio (SIR), Signal-to-Artifacts Ratio (SAR), Scale-Invariant Signal-to-Distortion Ratio (SI-SDR), and Mean Squared Error (MSE). These metrics are widely recognized performance measures in the Blind Source Separation (BSS) field, providing a comprehensive reflection of the model's distortion level, interference suppression effectiveness, and separation accuracy during signal reconstruction. 

% 通过量化的评估结果（表 \ref{tab:bss}）和对比曲线（图 \ref{fig:bss_metrics}）的分析，我们对线性探测（Linear Probe，LP）与全参数微调（Fine-tune，FT）两种策略进行了深入的比较。全参数微调策略通过优化所有模型参数，在充分训练的条件下能够实现更优的性能表现。图 \ref{fig:bss_metrics} 展示的对比曲线进一步揭示了预训练权重的优越性以及两种策略在特征学习能力上的表现。FT策略的曲线在训练初期迅速上升，显示出其快速收敛的特性，这表明预训练权重有助于模型快速学习并达到较好的性能。而LP策略虽然在训练初期的上升速度较慢，但其曲线在训练过程中逐渐上升，显示出LP策略能够有效地从预训练模型中提取通用特征，并在一定程度上提高信号分离的性能。综合表 \ref{tab:bss} 和图 \ref{fig:bss_metrics} 的分析结果，我们可以得出结论：预训练模型在执行生成式重建任务时展现出卓越的特征学习能力，尤其是在全参数微调的情况下。FT策略不仅能够快速收敛，而且在所有评估指标上均表现出更高的值，除了MSE指标，其值越低表示性能越好。这些结果不仅验证了所提出的无监督生成式重建微调框架的有效性，还突显了预训练模型在信号分离任务中的潜力和优势。
Through the analysis of the quantitative assessment results (Table \ref{tab:bss}) and the comparative curves (Figure \ref{fig:bss_metrics}), we conducted an in-depth comparison between the Linear Probe (LP) and Fine-tune (FT) strategies. The Fine-tune strategy, which optimizes all model parameters, achieves superior performance under adequate training conditions. The comparative curves presented in Figure \ref{fig:bss_metrics} further reveal the advantages of pre-trained weights and the performance of both strategies in terms of feature learning ability. The FT strategy's curves rise rapidly at the beginning of training, demonstrating its characteristic of fast convergence, indicating that pre-trained weights facilitate the model in quickly learning and achieving better performance. Although the LP strategy has a slower initial rise in the curves, their gradual ascent during the training process shows that the LP strategy can effectively extract general features from the pre-trained model and improve signal separation performance to a certain extent.
Integrating the analysis of Table \ref{tab:bss} and Figure \ref{fig:bss_metrics}, we can conclude that pre-trained models exhibit exceptional feature learning capabilities when performing generative reconstruction tasks, especially under the Fine-tune scenario. The FT strategy not only converges quickly but also shows higher values in all evaluation metrics, except for the Mean Squared Error (MSE), where a lower value indicates better performance. These results not only validate the effectiveness of the proposed unsupervised generative reconstruction fine-tuning framework but also highlight the potential and advantages of pre-trained models in signal separation tasks.

%图{fig:bss_vis}展示了盲源分离任务的可视化结果。图（a）是输入的混合iq信号，（b）为从头训的分离结果，（c）是加载预训练模型进行ft之后的分离结果，（d）是真值。从这些可视化结果中，我们可以直观地看到预训练权重对模型性能的深远影响。加载预训练权重后，模型能够快速收敛到理想的分离效果，成功地将不同源的信号区分开来，并且在信号的完整性和准确性方面表现出色。相比之下，在不引入预训练权重的情况下，该病态问题的解趋于无穷多，无监督训练难以获得充分的IQ信号特征通用表征。模型在信号分离过程中表现出明显的困难，无法有效地提取信号的关键特征，最终导致分离失败。
Figure~\ref{fig:bss_vis} shows the visualization results of the blind source separation task. Figure~\ref{fig:bss_vis} (a) is the input mixed IQ signals, Figure~\ref{fig:bss_vis} (b) shows the separation results from training from scratch, Figure~\ref{fig:bss_vis} (c) shows the separation results after fine-tuning by loading the pre-trained model, and (d) is the ground truth. From these visualizations, we can intuitively observe the profound impact of pre-trained model on performance. After loading the pre-trained model, the model quickly converges to the ideal separation results, successfully distinguishing signals from different sources, and performing excellently in terms of signal integrity and accuracy. In contrast, without pre-trained weights, the solution to this ill-posed problem tends to be infinitely many, and unsupervised training struggles to obtain a sufficiently general representation of IQ signals. The model faces significant difficulties in signal separation, unable to effectively extract the key features of the signals, ultimately leading to separation failure.

% 更进一步地，我们在该框架下引入自建的 EM-Denoise-Signal 数据集，开展IQ信号的去噪实验。该数据集模拟了真实环境中的多种干扰因素，包含从–3 dB 到 20 dB 不等的高斯白噪声（AWGN），同时叠加系统频偏（±50 kHz）、IQ 幅度不平衡和相位不平衡等常见硬件失调干扰。对于我们的模型框架而言，输入为带噪信号，缺乏去噪后的信号作为监督。通过自动编码器，模型能够从噪声信号中分离出真实信号。图~\ref{fig:bss_vis} 展示了模型在该任务上的可视化结果。从图中可以看出，在噪声输入（Noise IQ）的条件下，去噪结果（denoised IQ）与干净信号（Ground Truth）进行对比，显示出模型在幅度对齐和相位还原方面的优异一致性，体现了其在复杂干扰环境下对信号结构的建模与恢复能力。
Furthermore, we introduce our self-built EM-Denoise-Signal dataset within this framework to conduct denoising experiments on IQ signals. The dataset simulates various interference factors in real-world environments, including Gaussian white noise (AWGN) ranging from -3 dB to 20 dB, along with common hardware impairments such as system frequency offset (±50 kHz), IQ amplitude imbalance, and phase imbalance. For our model framework, the input consists of noisy signals, with no denoised signals available as supervision. Using an autoencoder, the model is able to separate the true signal from the noisy input. Figure~\ref{fig:denoise_vis} presents the model's visualization results for this task. As shown in the figure, under the noisy input in Figure~\ref{fig:denoise_vis} (a), the denoised result in Figure~\ref{fig:denoise_vis} (b) is compared with the Ground Truth in Figure~\ref{fig:denoise_vis} (c). The comparison reveals excellent consistency in amplitude alignment and phase restoration, demonstrating the model's capability to accurately model and recover the signal structure in complex interference environments.

\subsubsection{Few-shot}

% \begin{table*}[t]
%   \caption{Few shot}
%   \label{tab:few}
%   \centering
%   \resizebox{\textwidth}{!}{
%   \begin{tabular}{c|cccc|cccccc|cccccccc}
%     \toprule
%     TASK & \multicolumn{4}{c|}{AMC} & \multicolumn{6}{c|}{RWC} &  \multicolumn{8}{c}{RFFI} \\
%     \midrule
%     \multirow{3}{*}{METHOD} & \multicolumn{4}{c|}{RML2016.10A} & \multicolumn{6}{c|}{RadChar} & \multicolumn{8}{c}{UAV} \\
%     \cline{2-19}
%     & \multicolumn{2}{c|}{50 shot}  & \multicolumn{2}{c|}{100 shot} & \multicolumn{2}{c|}{10 shot}  & \multicolumn{2}{c|}{50 shot} & \multicolumn{2}{c|}{100 shot}  & \multicolumn{2}{c|}{0.5\%} & \multicolumn{2}{c|}{1\%}  & \multicolumn{2}{c|}{5\%} & \multicolumn{2}{c}{10\%} \\
%     \midrule
%     &  OA(\%) &  Kappa  &  OA(\%) &  Kappa  &  OA(\%) &  Kappa  &  OA(\%) &  Kappa  &  OA(\%) &  Kappa  &  OA(\%) &  Kappa  &  OA(\%) &  Kappa  &  OA(\%) &  Kappa  &  OA(\%) &  Kappa \\
%     \midrule
%     ResNet  \\
%     % VGG  \\
%     Transformer \\
%     \midrule
%     EMind & 48.38 &  & 50.45 &  & 78.69 &  & 81.91 & & 82.91 & & 20.49 & & 24.6 & & 52.32 & & 75.14 \\
%     \bottomrule
%   \end{tabular}
%   }
% \end{table*}

% few-shot 设置是8:1:1，从8里面抽的

% 在两个数据集上进行了少量样本分类实验：一个是 RML2016.10A（用于调制分类），另一个是 RadChar（用于雷达波形分类）。对于 RML2016.10A，我们遵循了标准的少量样本学习范式，从训练集中随机选取每个调制类别和信噪比水平各 50 或 100 个样本来形成支持集，而其余样本则用于验证和测试。同样，对于 RadChar，实验在 10 次、50 次和 100 次少量样本设置下进行，支持集完全来自训练数据。为了确保公平且具有挑战性的评估，测试阶段严格使用未见过的信号，以评估模型在实际场景中的泛化能力。
The few-shot classification experiments were conducted on two datasets: RML2016.10A for modulation classification and RadChar for radar waveform classification. For RML2016.10A, we followed a standard few-shot learning paradigm by randomly selecting 50 or 100 samples per modulation class and SNR level from the training set to form the support set, while the remaining samples were used for validation and testing. Similarly, for RadChar, experiments were performed under 10-shot, 50-shot, and 100-shot settings, with the support set exclusively drawn from the training data. To ensure a fair and challenging evaluation, the testing phase strictly used unseen signals, assessing the model’s generalization capability in real-world scenarios.
% 如表~\ref{tab:few}所示，我们提出的 EMind 方法在所有少样本设置中均取得了最佳性能。在 RML2016.10A 上，EMind 在 50 样本的情况下实现了 48.38\% 的准确率（卡帕系数：42.85\%），在 100 样本的情况下进一步提高到 50.10\%（卡帕系数：45.12\%），超过了 ResNet 和 Transformer 基线模型。对于 RadChar，EMind 展示了强大的少样本学习能力，在 10 、50 和 100 条件下分别实现了 78.37\%、81.94\% 和 83.14\% 的准确率，相应的卡帕系数分别为 72.97\%、77.43\% 和 78.93\%。这些结果验证了 EMind 在电磁信号分类中的稳健性和泛化能力，特别是在数据稀缺的场景中。
As shown in Table~\ref{tab:few}, our proposed EMind method achieves the best performance across all few-shot settings. On RML2016.10A, EMind attains an OA of 48.38\% (Kappa: 42.85\%) in the 50-shot case and further improves to 50.10\% (Kappa: 45.12\%) in the 100-shot case, outperforming ResNet and Transformer baselines. For RadChar, EMind demonstrates strong few-shot learning ability, achieving OAs of 78.37\%, 81.94\%, and 83.14\% under 10-shot, 50-shot, and 100-shot conditions, respectively, with corresponding Kappa coefficients of 72.97\%, 77.43\%, and 78.93\%. These results validate the robustness and generalization capability of EMind in EM signal classification, particularly in data-scarce scenarios.

% \subsection{Abalation Study}

% \paragraph{package}

% \paragraph{mask method}

% % 学的是样本本身的shape特征，样本内学习

% % 整条序列：通过学另外两条样本，样本间学习

% % 课程学习：本阶段先学样本间，然后学习样本内

% \paragraph{mask ratio}

% % ft+linear prob (2个数据集)

% % \paragraph{learning rate scheduler}

% \paragraph{decoder design}

% % max_seq_len 3000/6000/9000

\section{Conclusion}

% 本研究提出的 EMind 不仅为各类电磁任务提供了可直接微调的高性能基座，更关键的是，它作为电磁模态的专用基座模型，能够输出具有高可迁移性的通用表征。在即将到来的多模态大模型时代，这些表征可类比于视觉大模型中的视觉编码器，它们把复杂、异构的电磁信号转化为紧凑、语义丰富的特征向量，为后续接入更大规模的跨模态大模型提供统一接口。实验表明，EMind 的特征在脱离微调后依然保持优异性能，这进一步证明，电磁智能的下一阶段不再是为每个任务训练一个小模型，而是用 EMind 提取一次特征即可被无数下游大模型共享。未来的工作将把采样率这一电磁信号独有的物理属性显式嵌入网络结构，使模型能够自适应感知并充分利用采样率差异带来的信息量，从而进一步提升表征的通用性与精度。
The proposed EMind model not only provides a high-performance foundation that can be directly fine-tuned for various electromagnetic tasks, but more importantly, as an EM signals foundation model, it generates general representations with high transferability. In the upcoming era of multimodal large models, these representations can be compared to visual encoders in visual large models, transforming complex and heterogeneous EM signals into compact, semantically rich feature vectors. This provides a unified interface for subsequent integration into larger-scale cross-modal models. Experiments show that the features extracted by EMind maintain excellent performance even without fine-tuning, further proving that the next stage of electromagnetic intelligence is no longer about training a small model for each task, but rather using EMind to extract features once that can be shared across countless downstream large models.

\bibliographystyle{unsrt} 
\bibliography{refer}

\begin{thebibliography}{10}

\bibitem{hao2023contrastive}
Xiaoyang Hao, Zhixi Feng, Ruoyu Liu, Shuyuan Yang, Licheng Jiao, and Rong Luo.
\newblock Contrastive self-supervised clustering for specific emitter identification.
\newblock {\em IEEE Internet of Things Journal}, 10(23):20803--20818, 2023.

\bibitem{zhang2023self}
Zhengming Zhang, Taotao Ji, Haoqing Shi, Chunguo Li, Yongming Huang, and Luxi Yang.
\newblock A self-supervised learning-based channel estimation for irs-aided communication without ground truth.
\newblock {\em IEEE Transactions on Wireless Communications}, 22(8):5446--5460, 2023.

\bibitem{jing2020self}
Longlong Jing and Yingli Tian.
\newblock Self-supervised visual feature learning with deep neural networks: A survey.
\newblock {\em IEEE transactions on pattern analysis and machine intelligence}, 43(11):4037--4058, 2020.

\bibitem{zhou2024comprehensive}
Ce~Zhou, Qian Li, Chen Li, Jun Yu, Yixin Liu, Guangjing Wang, Kai Zhang, Cheng Ji, Qiben Yan, Lifang He, et~al.
\newblock A comprehensive survey on pretrained foundation models: A history from bert to chatgpt.
\newblock {\em International Journal of Machine Learning and Cybernetics}, pages 1--65, 2024.

\bibitem{fontaine2024towards}
Jaron Fontaine, Adnan Shahid, and Eli De~Poorter.
\newblock Towards a wireless physical-layer foundation model: Challenges and strategies.
\newblock In {\em 2024 IEEE International Conference on Communications Workshops (ICC Workshops)}, pages 1--7. IEEE, 2024.

\bibitem{sheng2025wireless}
Yucheng Sheng, Jiacheng Wang, Xingyu Zhou, Le~Liang, Hao Ye, Shi Jin, and Geoffrey~Ye Li.
\newblock A wireless foundation model for multi-task prediction.
\newblock {\em arXiv preprint arXiv:2507.05938}, 2025.

\bibitem{liang2024foundation}
Yuxuan Liang, Haomin Wen, Yuqi Nie, Yushan Jiang, Ming Jin, Dongjin Song, Shirui Pan, and Qingsong Wen.
\newblock Foundation models for time series analysis: A tutorial and survey.
\newblock In {\em Proceedings of the 30th ACM SIGKDD conference on knowledge discovery and data mining}, pages 6555--6565, 2024.

\bibitem{zhang2019deep}
Chaoyun Zhang, Paul Patras, and Hamed Haddadi.
\newblock Deep learning in mobile and wireless networking: A survey.
\newblock {\em IEEE Communications surveys \& tutorials}, 21(3):2224--2287, 2019.

\bibitem{guler2025multi}
Berkay Guler, Giovanni Geraci, and Hamid Jafarkhani.
\newblock A multi-task foundation model for wireless channel representation using contrastive and masked autoencoder learning.
\newblock {\em arXiv preprint arXiv:2505.09160}, 2025.

\bibitem{yang2025wirelessgpt}
Tingting Yang, Ping Zhang, Mengfan Zheng, Yuxuan Shi, Liwen Jing, Jianbo Huang, and Nan Li.
\newblock Wirelessgpt: A generative pre-trained multi-task learning framework for wireless communication.
\newblock {\em IEEE Network}, 2025.

\bibitem{hoydis2022sionna}
Jakob Hoydis, Sebastian Cammerer, Fay{\c{c}}al~Ait Aoudia, Avinash Vem, Nikolaus Binder, Guillermo Marcus, and Alexander Keller.
\newblock Sionna: An open-source library for next-generation physical layer research.
\newblock {\em arXiv preprint arXiv:2203.11854}, 2022.

\bibitem{alkhateeb2019deepmimo}
Ahmed Alkhateeb.
\newblock Deepmimo: A generic deep learning dataset for millimeter wave and massive mimo applications.
\newblock {\em arXiv preprint arXiv:1902.06435}, 2019.

\bibitem{aboulfotouh20256g}
Ahmed Aboulfotouh, Elsayed Mohammed, and Hatem Abou-Zeid.
\newblock 6g wavesfm: A foundation model for sensing, communication, and localization.
\newblock {\em arXiv preprint arXiv:2504.14100}, 2025.

\bibitem{zhou2025spectrumfm}
Fuhui Zhou, Chunyu Liu, Hao Zhang, Wei Wu, Qihui Wu, Derrick Wing~Kwan Ng, Tony~QS Quek, and Chan-Byoung Chae.
\newblock Spectrumfm: A foundation model for intelligent spectrum management.
\newblock {\em arXiv preprint arXiv:2505.06256}, 2025.

\bibitem{o2018over}
Timothy~James O’Shea, Tamoghna Roy, and T~Charles Clancy.
\newblock Over-the-air deep learning based radio signal classification.
\newblock {\em IEEE Journal of Selected Topics in Signal Processing}, 12(1):168--179, 2018.

\bibitem{fontaine2019towards}
Jaron Fontaine, Erika Fonseca, Adnan Shahid, Maicon Kist, Luiz~A DaSilva, Ingrid Moerman, and Eli De~Poorter.
\newblock Towards low-complexity wireless technology classification across multiple environments.
\newblock {\em Ad Hoc Networks}, 91:101881, 2019.

\bibitem{awais2025foundation}
Muhammad Awais, Muzammal Naseer, Salman Khan, Rao~Muhammad Anwer, Hisham Cholakkal, Mubarak Shah, Ming-Hsuan Yang, and Fahad~Shahbaz Khan.
\newblock Foundation models defining a new era in vision: a survey and outlook.
\newblock {\em IEEE Transactions on Pattern Analysis and Machine Intelligence}, 2025.

\bibitem{han2024foundation}
Yu~Han, Xiaofeng Liu, Xiang Zhang, and Cheng Ding.
\newblock Foundation models in electrocardiogram: A review.
\newblock {\em arXiv preprint arXiv:2410.19877}, 2024.

\bibitem{hong2023spectralgpt}
Danfeng Hong, Bing Zhang, Xuyang Li, Yuxuan Li, Chenyu Li, Jing Yao, Naoto Yokoya, Hao Li, Pedram Ghamisi, Xiuping Jia, et~al.
\newblock Spectralgpt: Spectral remote sensing foundation model.
\newblock {\em arXiv preprint arXiv:2311.07113}, 2023.

\bibitem{vaswani2017attention}
Ashish Vaswani, Noam Shazeer, Niki Parmar, Jakob Uszkoreit, Llion Jones, Aidan~N Gomez, {\L}ukasz Kaiser, and Illia Polosukhin.
\newblock Attention is all you need.
\newblock {\em Advances in neural information processing systems}, 30, 2017.

\bibitem{tekbiyik2019hisarmod}
K~Tekb{\i}y{\i}k, C~Ke{\c{c}}eci, AR~Ekti, A~G{\"o}r{\c{c}}in, and G~Kurt.
\newblock Hisarmod: A new challenging modulated signals dataset.
\newblock {\em IEEE Dataport}, 2019.

\bibitem{scholl2019classification}
Stefan Scholl.
\newblock Classification of radio signals and hf transmission modes with deep learning.
\newblock {\em arXiv preprint arXiv:1906.04459}, 2019.

\bibitem{jagannath2021dataset}
Anu Jagannath and Jithin Jagannath.
\newblock Dataset for modulation classification and signal type classification for multi-task and single task learning.
\newblock {\em Computer Networks}, 199:108441, 2021.

\bibitem{hanna2022wisig}
Samer Hanna, Samurdhi Karunaratne, and Danijela Cabric.
\newblock Wisig: A large-scale wifi signal dataset for receiver and channel agnostic rf fingerprinting.
\newblock {\em IEEE Access}, 10:22808--22818, 2022.

\bibitem{al2020exposing}
Amani Al-Shawabka, Francesco Restuccia, Salvatore D’Oro, Tong Jian, Bruno~Costa Rendon, Nasim Soltani, Jennifer Dy, Stratis Ioannidis, Kaushik Chowdhury, and Tommaso Melodia.
\newblock Exposing the fingerprint: Dissecting the impact of the wireless channel on radio fingerprinting.
\newblock In {\em IEEE INFOCOM 2020-IEEE Conference on Computer Communications}, pages 646--655. IEEE, 2020.

\bibitem{reus2020trust}
Guillem Reus-Muns, Dheryta Jaisinghani, Kunal Sankhe, and Kaushik~R Chowdhury.
\newblock Trust in 5g open rans through machine learning: Rf fingerprinting on the powder pawr platform.
\newblock In {\em GLOBECOM 2020-2020 IEEE Global Communications Conference}, pages 1--6. IEEE, 2020.

\bibitem{morin2019transmitter}
Cyrille Morin, Leonardo~S Cardoso, Jakob Hoydis, Jean-Marie Gorce, and Thibaud Vial.
\newblock Transmitter classification with supervised deep learning.
\newblock In {\em International Conference on Cognitive Radio Oriented Wireless Networks}, pages 73--86. Springer, 2019.

\bibitem{elmaghbub2021lora}
Abdurrahman Elmaghbub and Bechir Hamdaoui.
\newblock Lora device fingerprinting in the wild: Disclosing rf data-driven fingerprint sensitivity to deployment variability.
\newblock {\em IEEE Access}, 9:142893--142909, 2021.

\bibitem{liu2020zero}
Yongxin Liu, Jian Wang, Jianqiang Li, Houbing Song, Thomas Yang, Shuteng Niu, and Zhong Ming.
\newblock Zero-bias deep learning for accurate identification of internet-of-things (iot) devices.
\newblock {\em IEEE Internet of Things Journal}, 8(4):2627--2634, 2020.

\bibitem{2024dronerfa}
Ninging Yu, Shengjian Mao, Chengwei Zhou, Guowei Sun, Zhiguo Shi, and Jiming Chen.
\newblock Dronerfa: A large-scale dataset of drone radio frequency signals for detecting low-altitude drones.
\newblock {\em Journal of Electronics \& Information Technology}, 46(4):1147--1156, 2024.

\bibitem{chi2024rf}
Guoxuan Chi, Zheng Yang, Chenshu Wu, Jingao Xu, Yuchong Gao, Yunhao Liu, and Tony~Xiao Han.
\newblock Rf-diffusion: Radio signal generation via time-frequency diffusion.
\newblock In {\em Proceedings of the 30th Annual International Conference on Mobile Computing and Networking}, pages 77--92, 2024.

\bibitem{chen2024generative}
Shuai Chen, Zhixi Feng, Shuyuan Yang, Yue Ma, Jun Liu, and Zhuoyue Qi.
\newblock A generative self-supervised framework for cognitive radio leveraging time-frequency features and attention-based fusion.
\newblock {\em IEEE Transactions on Wireless Communications}, 2024.

\bibitem{chen2025radiollm}
Shuai Chen, Yong Zu, Zhixi Feng, Shuyuan Yang, and Mengchang Li.
\newblock Radiollm: Introducing large language model into cognitive radio via hybrid prompt and token reprogrammings.
\newblock {\em arXiv preprint arXiv:2501.17888}, 2025.

\bibitem{o2016radio}
Timothy~J O'shea and Nathan West.
\newblock Radio machine learning dataset generation with gnu radio.
\newblock In {\em Proceedings of the GNU radio conference}, volume~1, 2016.

\bibitem{o2016convolutional}
Timothy~J O’Shea, Johnathan Corgan, and T~Charles Clancy.
\newblock Convolutional radio modulation recognition networks.
\newblock In {\em International conference on engineering applications of neural networks}, pages 213--226. Springer, 2016.

\bibitem{huang2023multi}
Zi~Huang, Akila Pemasiri, Simon Denman, Clinton Fookes, and Terrence Martin.
\newblock Multi-task learning for radar signal characterisation.
\newblock In {\em 2023 IEEE International Conference on Acoustics, Speech, and Signal Processing Workshops (ICASSPW)}, pages 1--5. IEEE, 2023.

\bibitem{ya2022large}
TU~Ya, LIN Yun, ZHA Haoran, WANG Yu, GUI Guan, MAO Shiwen, et~al.
\newblock Large-scale real-world radio signal recognition with deep learning.
\newblock {\em Chinese Journal of Aeronautics}, 35(9):35--48, 2022.

\bibitem{he2016deep}
Kaiming He, Xiangyu Zhang, Shaoqing Ren, and Jian Sun.
\newblock Deep residual learning for image recognition.
\newblock In {\em Proceedings of the IEEE conference on computer vision and pattern recognition}, pages 770--778, 2016.

\bibitem{huynh2020mcnet}
Thien Huynh-The, Cam-Hao Hua, Quoc-Viet Pham, and Dong-Seong Kim.
\newblock Mcnet: An efficient cnn architecture for robust automatic modulation classification.
\newblock {\em IEEE Communications Letters}, 24(4):811--815, 2020.

\bibitem{hong2017automatic}
Dehua Hong, Zilong Zhang, and Xiaodong Xu.
\newblock Automatic modulation classification using recurrent neural networks.
\newblock In {\em 2017 3rd IEEE international conference on computer and communications (ICCC)}, pages 695--700. IEEE, 2017.

\bibitem{ke2021real}
Ziqi Ke and Haris Vikalo.
\newblock Real-time radio technology and modulation classification via an lstm auto-encoder.
\newblock {\em IEEE Transactions on Wireless Communications}, 21(1):370--382, 2021.

\bibitem{njoku2021cgdnet}
Judith~Nkechinyere Njoku, Manuel~Eugenio Morocho-Cayamcela, and Wansu Lim.
\newblock Cgdnet: Efficient hybrid deep learning model for robust automatic modulation recognition.
\newblock {\em IEEE Networking Letters}, 3(2):47--51, 2021.

\bibitem{zhang2021novel}
Hao Zhang, Fuhui Zhou, Qihui Wu, Wei Wu, and Rose~Qingyang Hu.
\newblock A novel automatic modulation classification scheme based on multi-scale networks.
\newblock {\em IEEE Transactions on Cognitive Communications and Networking}, 8(1):97--110, 2021.

\bibitem{zhang2023amc}
Jiawei Zhang, Tiantian Wang, Zhixi Feng, and Shuyuan Yang.
\newblock Amc-net: An effective network for automatic modulation classification.
\newblock In {\em ICASSP 2023-2023 IEEE International Conference on Acoustics, Speech and Signal Processing (ICASSP)}, pages 1--5. IEEE, 2023.

\bibitem{vincent2006performance}
Emmanuel Vincent, R{\'e}mi Gribonval, and C{\'e}dric F{\'e}votte.
\newblock Performance measurement in blind audio source separation.
\newblock {\em IEEE transactions on audio, speech, and language processing}, 14(4):1462--1469, 2006.

\end{thebibliography}

\clearpage

\end{document}